\documentclass[11pt,a4paper]{article}
\pdfoutput=1
\usepackage[english]{babel}
\usepackage[utf8]{inputenc}
\usepackage{jheppub}
\usepackage{nccmath}
\usepackage{amsfonts}
\usepackage{nameref}
\usepackage{latexsym}
\usepackage{appendix} 
\usepackage{amssymb,amsmath} 
\usepackage{amsthm}
\usepackage{mathtools}
\usepackage{epstopdf}
\usepackage{verbatim}
\usepackage{float}
\usepackage{makecell}
\usepackage{hyperref}
\usepackage{multirow} 
\usepackage{graphicx}
\usepackage{pdfpages}
\usepackage{dcolumn}
\usepackage{bm}
\usepackage{placeins}
\usepackage{systeme}

\DeclareMathOperator{\mm}{\mathcal{M}}
\numberwithin{equation}{section}
\allowdisplaybreaks

\title{Little String Theories on Curved Manifolds}
\author{Ofer Aharony,}
\author{Mikhail Evtikhiev}
\author{and Andrey Feldman}
\affiliation{Department of Particle Physics and Astrophysics, \\
Weizmann Institute of Science, Rehovot 7610001, Israel}
\emailAdd{ofer.aharony@weizmann.ac.il}
\emailAdd{mikhail.evtikhiev@weizmann.ac.il}
\emailAdd{andrey.feldman@weizmann.ac.il}
\abstract{In this paper, we study the 6d Little String Theory (LST) (the decoupled theory on the worldvolume of $N$ NS5-branes) on curved manifolds, by using its holographic duality to Type II string theory in asymptotically linear dilaton backgrounds. We focus on backgrounds with a large number of Killing vectors (namely, products of maximally symmetric spaces), without requiring supersymmetry (we do not turn on any background fields except the metric). LST is non-local so it is not obvious which spaces it can be defined on; we show that holography implies that the theory cannot be put on negatively curved spaces, but only on spaces with zero or positive curvature. For example, one cannot put LST on a product of an anti-de Sitter space times another space, without turning on extra background fields. On spaces with positive curvature, such as $S^6$, $\mathbb{R}^2\times S^4$, $S^3\times S^3$, etc., we typically find (for large $N$) dual holographic backgrounds which are weakly coupled and weakly curved everywhere, so that they can be well-described by Type II supergravity. In some cases more than one smooth solution exists for LST on the same space, and they all contribute to the partition function. We also study the thermodynamical properties of LST compactified on spheres, finding the leading correction to the Hagedorn behavior of the spectrum, which is different on curved space than on flat space. We discuss the holographic renormalization procedure, which must be implemented in order to get a finite free energy for the LST; we do not know how to implement it for general spaces, but we can (and we do) implement it for the theory compactified on $S^4$.}

\begin{document}


\maketitle

\flushbottom

\selectlanguage{english}

\section{Introduction and summary} \label{sec:intro}

Little string theory (LST) is a mysterious interacting non-local non-gravitational theory, which was originally discovered as a decoupled effective worldvolume theory on a flat stack of $N$ NS5-branes ($N>1$) in Type II string theory \cite{BerkoozRozaliSeiberg, Seiberg} (see \cite{OferReview, KutasovReview} for reviews). The difference between the Type IIA and IIB cases is that in the theory of Type IIA NS5-branes, they preserve a chiral half of the supersymmetry (SUSY), so the resulting LST has 6d $\mathcal{N} = (2, 0)$ SUSY, while in the Type IIB case, it has $\mathcal{N} = (1, 1)$ SUSY. The LST on the branes can be decoupled from the rest of the string theory dynamics by sending the asymptotic string coupling $g_s$ to zero, while keeping $E/M_s$ fixed, where $E$ is a typical energy we work with, and $M_s$ is the string scale, which is the only parameter of the theory (for a given value of $N$). LST is in many ways between a quantum field theory (QFT) and a string theory --- it is non-local\footnote{In particular, when compactified on a circle, it enjoys a T-duality relating the $\mathcal{N} = (2,0)$ LST to the $\mathcal{N} = (1,1)$ one. This duality descends to 6d LST from the full 10d string theory. Similarly, the LST compactified on $T^d$ has an $\mathrm{SO} (d, d, \mathbb{Z})$ T-duality symmetry.} and has a Hagedorn density of states at high energies,
but it doesn't contain gravity and it has well-defined off-shell Green's functions, like QFT \cite{AharonyGiveonKutasov}, so the proper understanding of LST can hopefully improve our understanding of both subjects. The scale $M_s$ appears, for instance, in the tension of BPS-saturated strings in the theory. At energies below this scale, the $\mathcal{N} = (2,0)$ LST flows to the 6d $\mathcal{N} = (2,0)$ superconformal field theories (SCFTs), while the $\mathcal{N} = (1,1)$ LST flows to an $\mathrm{SU}(N)$ maximally supersymmetric Yang-Mills (SYM) theory.


LST in flat space doesn't have any dimensionless parameters, which could have been used for a perturbative expansion. The best method for performing computations in LST is its holographic description \cite{AharonyBerkoozKutasovSeiberg} in terms of Type II string theory in an asymptotically linear dilaton background.\footnote{One other approach is via discrete light-cone quantization (DLCQ) \cite{AharonyBerkoozKachruSeibergSilverstein, WittenHiggsBranch}.} The background dual to the LST related to $N$ NS5-branes on $\mathbb{R}^{5,1}$ is given, in the string frame, by
\begin{equation}\label{flatholo}
\mathrm{d} s_{str}^2 = \mathrm{d} y_{\mu}^2 + N \alpha' \left( \mathrm{d} r^2 + \mathrm{d} \Omega_3^2 \right), \qquad g_s = \mathrm{e}^{-r},
\end{equation}
where $\mathrm{d} y_{\mu}^2$ is the metric on $\mathbb{R}^{5,1}$, $\alpha'$ is proportional to the inverse string tension, $\mathrm{d} \Omega_3^2$ is the metric on an $S^3$ of unit radius, $g_s$ is the string coupling, and there are $N$ units of 3-form flux on the $S^3$. At $r\to \infty$ the string coupling goes to zero, related to the decoupling from gravity, and we can turn on sources for the LST there. As $r\to -\infty$ the string coupling diverges so this description is not useful; however, the description is useful away from this region, and
there are various setups (such as \cite{GiveonKutasov}) where this strong coupling region is absent so that the holographic description is complete\footnote{In some examples the strongly coupled region may be controlled by lifting it from Type IIA to M-theory; for instance this is the case for the $\mathcal{N} = (2,0)$ LST on $\mathbb{R}^{5,1}$, whose dual description interpolates between a linear dilaton background of Type IIA string theory, and M-theory on $AdS_7\times S^4$.}  (we will find some additional examples of such setups in this paper). The fact that \eqref{flatholo} is a solution to string theory is clear from the worldsheet point of view, since the worldsheet theory is just the product of several SCFTs: free scalars for $\mathbb{R}^{5,1}$, a linear dilaton for the $r$ direction, an $\mathrm{SU}(2)_N$ Wess-Zumino-Witten (WZW) model for the $S^3$ factor, and their superpartners.

Naively, one may think that in order to construct a holographic description for LSTs on some curved space ${\cal M}$, one has to replace the $\mathrm{d} y_{\mu}^2$ factor in the metric \eqref{flatholo} by the metric of ${\cal M}$ as $r\to \infty$, and to look for appropriate solutions of string theory with that asymptotic behavior. Such a replacement works when ${\cal M}$ is Ricci-flat, corresponding to a conformal field theory (CFT) on the worldsheet,\footnote{At least at leading order in the $\alpha'$ expansion, and one can then systematically incorporate higher order corrections.} but it fails when the corresponding sigma-model on the worldsheet has a non-zero beta function. However, similar solutions do exist in which the metric on ${\cal M}$ depends on the radial direction (even as $r\to \infty$), as may be expected based on interpreting this direction as a scale on the worldsheet. Such solutions were found for supersymmetric compactifications of LST in \cite{MaldacenaNunez,MaldacenaNastase,BobevBomansGautason}, and for some specific non-supersymmetric curved backgrounds in \cite{Buchel}, and in this paper we study them in more generality.

If one puts LST on a compact manifold and flows to the IR, the resulting theory may be related to interesting strongly coupled systems. For example, the supersymmetric $S^2$ compactification leads in a specific limit of its dimensionless parameter to the 4d $\mathcal{N} = 1$ SYM theory \cite{MaldacenaNunez}, and the supersymmetric $S^3$ compactification leads in a similar limit to the 3d $\mathcal{N} = 1$ Yang-Mills-Chern-Simons theory \cite{MaldacenaNastase}.\footnote{In both cases supergravity breaks down in the corresponding limit of the holographic description \cite{NonAdS}, but for other values of the dimensionless parameter the holographic description is weakly coupled and weakly curved.} Similar relations should exist also in the non-supersymmetric cases we study in this paper. In order to preserve SUSY, one has to twist the theory in the compact directions, by adding background fields that effectively mix the Lorentz symmetry with the R-symmetry to allow for covariantly constant spinors on the compact manifold \cite{VafaTopological}. On the gravity side of the duality, this corresponds to switching on additional $p$-form fields
in addition to the 
dilaton $\Phi$ and the metric $G_{\mu \nu}$, which are turned on in flat space.

In this paper, we study non-supersymmetric LST on maximally symmetric manifolds and their products $\mathcal{M}$, without any additional background fields, using holography, by finding supergravity (SUGRA) solutions with a linear dilaton at infinity, of asymptotic topology $\mathcal{M} \times S^3 \times \mathbb{R}_{\Phi}$.
We begin in section \ref{setup} by describing the setup and the equations we solve, and in section \ref{solving} we discuss the solutions and how we obtain them.
In general we cannot construct such solutions analytically, but we can find them numerically. We can always find analytically the asymptotic form of the solutions, which is enough to identify for which spacetimes solutions exist, and to count their parameters.
Since LST is non-local and non-Lagrangian, it is a non-trivial question, on which manifolds $\mathcal{M}$ it can live. The asymptotic analysis shows that there are no appropriate solutions to SUGRA if $\mathcal{M} = AdS_6$ or if it contains $AdS_{q<6}$ as a factor. For products of spherical (or de Sitter) and flat factors, generically smooth solutions exist, with a circle or a sphere shrinking smoothly at some value $r=0$ of the radial coordinate, and with a finite string coupling everywhere. The parameters of these solutions may be taken to be the volumes (in string units) of the compact factors of space (the radius of curvature for de Sitter factors), evaluated in the asymptotic region (say, where the string coupling takes some specific very small value). One of the parameters always affects the solutions only by an overall shift of the dilaton (a rescaling of the string coupling), such that it does not affect the classical supergravity solutions we find, but only the loop corrections to them (which we do not discuss). The other parameters do affect the classical solutions.


There is one case where we have a smooth analytic solution, which is the case of LST on $\mathbb{R}^2\times S^4$ (the two flat directions can also be compactified on circles). In all other cases our solutions are numerical (there are a few cases where we have singular analytic solutions). In general, more than one solution may exist for LST on the same manifold (with the same parameters); these could have different topologies (with different geometrical factors shrinking to zero in the interior), or several solutions may exist with the same topology (this arises generically in the vicinity of singular solutions, as we discuss at the end of the paper in section \ref{oscs}). In such cases it is interesting to compute the action of the corresponding solutions, and to find the solution of minimal (Euclidean) action; when one of the directions of the LST is an $S^1$, we can view this circle as a Euclidean thermal circle corresponding to the canonical ensemble, and then this action is related to the thermal free energy. As usual in holography, computing the action requires holographic renormalization \cite{HenningsonSkenderis1,HenningsonSkenderis2,deHaroSolodukhinSkenderis,BianchiFreedmanSkenderis1,BianchiFreedmanSkenderis2,BalasubramanianKraus,Myers,EmparanJohnsonMyers,Mann,KrausLarsenSiebelink} (see \cite{CotronePonsTalavera, MarolfVirmani} for a discussion of holographic renormalization in the LST context, and \cite{Skenderis} for a review), and we discuss in section \ref{renorm} how to perform this procedure in our case. Unfortunately, for LST on generic manifolds the form of the divergences in the action is complicated, and an infinite number of counterterms may be required to obtain a finite action, so we are not able to compute the renormalized action. However, in special cases for which the asymptotic solution is particularly simple, namely the cases where the curved factors are an $S^4$, or $S^2\times S^2$ of equal radii, it turns out that only two counterterms are required to cancel the divergences, and we can explicitly perform the holographic renormalization.


Solutions which include an $S^1$ factor may be interpreted as Euclidean black holes, describing the thermal ensemble at the corresponding temperature (related to the radius of the $S^1$), and their Wick rotations may be interpreted as thermal states in Lorentzian LST. The holographic description allows for a simple computation of the thermal equation of state of LST, either by computing the black hole entropy through the Bekenstein-Hawking formula, or by computing the free energy as described in the previous paragraph (the two computations give the same answer). LST in flat space is known to have a Hagedorn behavior \cite{MaldacenaStrominger}
\begin{equation} \label{thermo}
E =T_H S, \qquad T_H = \frac{M_s}{2\pi \sqrt{N}},
\end{equation}
with an exponential density of states at high energies (note that for $N>2$ the corresponding exponent is larger than the one describing highly excited strings in the holographic background \eqref{flatholo}). In flat space, equation \eqref{thermo} is exact in classical string theory, and receives one-loop corrections that were analyzed in \cite{KutasovSahakyan} and that lead to the Hagedorn temperature being a limiting temperature for the existence of the canonical ensemble. In curved space, we show (as first shown in \cite{Buchel:2001dg,Gubser:2001eg,BuchelInstability} for the Maldacena-Nu\~{n}ez solution) that corrections to the Hagedorn behavior (of a somewhat different form) arise already classically, and again these lead to the Hagedorn temperature being a limiting one. We analyze the case of LST on $S^4$ in detail in section \ref{sec:hagedorn}, because in this case we can explicitly compute the free energy. We expect a similar thermodynamic behavior (summarized in equation \eqref{Equationofstate2} below) for any compactification of LST on a positively curved space.


In this paper we construct various supergravity solutions, and it would be interesting to analyze their properties further. We have not been able to perform a holographic renormalization of the action for most of our solutions, and it would be interesting to understand how to do this, and to use it to analyze the phase structure of the corresponding compactifications of LST. Since our solutions are non-supersymmetric, it is interesting to ask if they are stable or not, by analyzing small fluctuations around them; this may be complicated since most of our solutions are only known numerically, but it should be possible. More generally, one can compute the spectrum of excitations in the backgrounds we find; we expect generic solutions to exhibit a discrete spectrum of excitations, and it would be interesting to understand its properties. In flat space the spectrum of excitations around \eqref{flatholo} contains a continuum, and it would be interesting to understand what becomes of this in curved space. We construct our solutions in supergravity, and it would be interesting to see if some of them (in particular, the cases where we have analytic solutions) may correspond to exact conformal field theories on the worldsheet, and, more generally, how they are affected by stringy corrections.

There are many possible generalizations of our analysis. One can consider LST on other spacetimes that are not products of maximally-symmetric spaces, at the cost of generally needing to solve partial differential equations and not just ordinary ones as in our case.\footnote{Even for maximally-symmetric spaces, we took the simplest possible ansatz for the solutions, and it is possible that additional solutions exist, which do not preserve the full symmetries of this ansatz.} One can add additional background fields beyond the metric; in particular it is interesting to add background fields that preserve some supersymmetry. A few supersymmetric compactifications of LST have been analyzed already, as mentioned above, but many more should be possible, and in particular it should be possible to find supersymmetric compactifications of LST on products of anti-de Sitter space and spheres, related to the near-horizon limits of NS5-branes on such spaces. We focused in this paper on the 6d maximally supersymmetric LST, but our methods are completely general and should apply to any LST with a holographic description (for instance, our $\mathbb{R}^2\times S^4$ solution may be directly used to give a solution for 4d LSTs on $S^4$).

\section{The holographic setup} \label{setup}

We will be using the holographic correspondence to study LST, and we will be working in the low-energy supergravity approximation to Type II string theory, without turning on any Ramond-Ramond fields. Thus, the starting point for us will be the bosonic sector of the 10d SUGRA action in the Einstein frame with vanishing Ramond-Ramond fields,
\begin{equation} \label{action}
	I = -\frac{1}{2 \varkappa^2} \int \mathrm{d}^{10} x \sqrt{G} \left( R- \frac{1}{2} \partial_{\mu} \Phi \partial^{\mu} \Phi - \frac{1}{12} \mathrm{e}^{- \Phi} H_{\mu \nu \rho} H^{\mu \nu \rho} \right),
\end{equation}
where $\varkappa$ is Newton's constant in ten dimensions, and $G = | \mathrm{det} (G_{\mu \nu}) |$.


We are interested in solutions corresponding to LST in curved space. We will assume for simplicity that the $S^3$ factor in the string frame metric remains intact; it is clear from the worldsheet that this always gives a consistent solution, since it is a decoupled CFT (though it may not be the most general solution). We will also assume that $H = 0$ in the non-$S^3$ directions, so that we only turn on the dilaton and the metric in these directions. For simplicity, we consider LSTs on 6d manifolds ${\cal M}$ that are a product of maximally symmetric spaces of dimensions $d_k$ with metrics $\mathrm{d} s_k^2$, which can be spheres, $dS$ or $AdS$ spaces, or flat spaces ($\mathbb{R}^d$ or $T^d$). It is then natural to consider, as in \cite{Buchel}, 
the following ansatz for the Einstein frame metric:
\begin{equation}
	\mathrm{d} s^2 = c_2^2(r) \mathrm{d} r^2 + c_3^2(r) \mathrm{d} \Omega_3^2 + \sum_k c_{k}^2(r) \mathrm{d} s_k^2,
\end{equation}
where $c_i$ are the warp factors, depending only on the radial coordinate $r$.
Plugging in a solution of this sort, with $N$ units of $H$ flux on the $S^3$ and $H=0$ otherwise, the action is given by 
\begin{equation} 
	I = -\frac{1}{2 \varkappa^2} \int \mathrm{d}^{10} x \sqrt{G} \left( R - 2 N^2 \frac{1}{c_3^6 g} - \frac{1}{2} \frac{g^{\prime 2}}{g^2 c_2^2} \right) \label{baction},
\end{equation}
where $g = \mathrm{e}^{\Phi}$ is the string coupling, and the derivatives are taken with respect to $r$. The resulting equations of motion are (with a specific convenient normalization of the $\mathrm{d} s_k^2$ factors)
\begin{align}
	& \left (\frac{g'\lambda}{g} \right)' + 2 N^2 \frac{\lambda c_2^2}{g c_3^6} = 0, \\
	& \left (\frac{c_3' \lambda}{c_3} \right)' - 2 \frac{\lambda c_2^2}{c_3^2} + \frac{3 N^2}{2} \frac{\lambda c_2^2}{g c_3^6} = 0, \\
	& \left (\frac{c_k' \lambda}{c_k} \right)' - \left( d_k - 1 \right) \kappa_k \frac{\lambda c_2^2}{c_k^2} - \frac{N^2}{2} \frac{\lambda c_2^2}{g c_3^6} = 0, \label{LSTeqn}
\end{align}
where $\lambda \equiv \prod_k c_k^{d_k} c_3^3/c_2$, and $\kappa_k$ is the sign of its curvature of $\mathrm{d}s_k^2$ ($1$ for the sphere and de Sitter, $0$ for $\mathbb{R}^p$ and $(-1)$ for anti-de Sitter). In the $dS_6$ case, the equations are the same as the ones in \cite{Buchel}. 
The function $c_2(r)$ can be fixed to any desired value by a diffeomorphism of $r$, so its equation of motion gives a constraint $C[c_i(r)]$ of first order in derivatives rather than a second order equation. This constraint does not follow from the other equations of motion, but it is obeyed on their solutions (namely, if we consider a solution to the other equations such that the initial conditions at some value of $r$ satisfy $C=0$, we'll have $C'=C=0$ for any value of $r$).  

As discussed above, we know from the worldsheet that a solution for $c_3$ is given by
\begin{equation}
c_3 = \sqrt{N} g^{-1/4},
\end{equation}
and it will be convenient to choose the radial coordinate such that also
\begin{equation}
c_2 = \sqrt{N} g^{-1/4},
\end{equation}
and the string frame metric in the radial direction is simply $(N \mathrm{d}r^2)$. The equations for the remaining warp factors simplify if we go to the string frame variables $\zeta_k$ (as in \cite{Buchel}) via
\begin{equation}
c_k = \sqrt{N} \zeta_k g^{-1/4}.
\end{equation}
The $\zeta_k$'s differ from the standard string frame by a factor of $\sqrt{N}$, such that we are measuring distances in units of $\sqrt{N \alpha'}$; with this scaling the classical equations of motion are independent of $N$. Any non-singular solution is then weakly curved for large $N$, and stringy curvature solutions may be ignored in that limit.

Defining $\lambda_0 = \prod\limits_k \zeta_k^{d_k}$, we find that the remaining equations can be written as
\begin{align} 
	& \left(\frac{g'\lambda_0}{g^3}\right)' + 2 \frac{\lambda_0}{g^2} = 0, \label{LSTstring0} \\
	& \left(\frac{\zeta'_k \lambda_0}{\zeta_k g^2}\right)' - (d_k - 1)\kappa_k  \frac{\lambda_0}{g^2 \zeta_k^2} = 0, \label{LSTstring} \\
	& \sum\limits_k \left[ \frac{d_k (d_k-1)}{4} \left(\frac{ \zeta_k^{\prime 2}}{\zeta_k^2} - \kappa_k \frac{1}{\zeta_k^2} \right)  - d_k \frac{\zeta_k' g'}{\zeta_k g} + \sum\limits_{l < k} \frac{d_k d_{l}}{2} \frac{\zeta'_k \zeta'_{l}}{\zeta_k \zeta_{l}} \right] + \frac{g^{\prime 2}}{g^2} = 1,
\end{align}
where the last equation is the constraint. 

From the worldsheet point of view, \eqref{LSTstring} captures the fact that while flat spaces ($\kappa_k=0$) give CFTs, curved spaces have a non-vanishing beta function on the worldsheet for their curvature, which has to be canceled in the context of our solution by having their size depend on the radial direction $r$ (and also modifying the linear dilaton solution).


\section{Solving the equations of motion}\label{solving}

The equations of motion \eqref{LSTstring0}, \eqref{LSTstring} that we have obtained are strongly nonlinear, so it is difficult to find analytic solutions. Indeed, we couldn't find analytic solutions for most cases (we will discuss the exceptions later), but in many cases it is possible to solve the equations numerically and to prove the existence of a smooth solution. 

We will begin in section \ref{sec:asymp} by analyzing the asymptotic solutions for large $r$. This will tell us which spacetimes are allowed for the LST, and will enable us to identify the parameters (the non-normalizable modes of the solutions) and the normalizable fluctuations.

We are interested in finding smooth solutions, in which there is some minimal radial position $r$ where space smoothly ends; without loss of generality we can always choose to shift the radial coordinate such that this happens at $r=0$. 
Clearly, for this to happen one of the compact factors in ${\cal M}$ has to shrink to zero there; this can be a flat $S^1$ factor, or a positively curved $S^d$ factor. We can then solve the equations of motion by numerically integrating them towards positive values of $r$, starting from the smoothly shrinking solution for one of the compact factors, with some finite initial size for all the other factors (we should also start with some fixed value for the dilaton, though the equations are independent of shifting the dilaton by multiplying $g$ by a constant, so any choice here gives the same solutions). The boundary conditions that we impose are thus
\begin{equation}
g(0)= \mathrm{const}, \quad g'(0) = 0
\end{equation}
for the string coupling,
\begin{equation}
\zeta_k(0)= 0, \quad \zeta_k'(0) = 1
\end{equation}
for the specific warp factor that vanishes at $r=0$, and
\begin{equation}
\zeta_k(0)= \mathrm{const}, \quad \zeta_k'(0) = 0
\end{equation}
for all other warp factors.

In practice, the method we will use to get a solution of the equations of motion is as follows:
\begin{itemize}
	\item[1.] {\bf Series piece of solution.} We start by expanding the warp factors and dilaton around $r=0$ in a power series in $r$. The free parameters are the non-vanishing warp factors $\zeta_k(0)$ there (without loss of generality we can choose the dilaton $g(0)=1$), and having a smooth solution determines all functions and their derivatives in terms of these warp factors. We can then analytically obtain the solution as a power series in $r$, obtaining some number of terms.
The resulting power series solution will have very good precision inside a certain finite radius of convergence, but will behave badly outside it. As the convergence radius grows very slowly with the number of terms, we choose some point inside the convergence area and go to the next step.
	\item[2.] {\bf Numerical piece of solution.} The next step is to solve \eqref{LSTstring0}, \eqref{LSTstring} numerically toward larger values of $r$, with the initial conditions provided by the results of the previous step. Solving the equations of motion numerically starting directly with the initial conditions at zero fails due to computational issues, but there is no need for it.
	\item[3.] {\bf Asymptotic piece of solution.} At a certain large $r$ the numerical solution will fail (because the dilaton value is too small to do a decent numerical computation). Before we reach this value, we sew our numerical solution with the asymptotic analytic solution mentioned above. 
\end{itemize}
This procedure leads to a  family of smooth solutions, with various values of the asymptotic parameters, which are functions of the initial warp factors at $r=0$.

\subsection{Asymptotic limit} \label{sec:asymp}

Let us start with the asymptotic limit. We say that the holographic dual we study admits an LST-like solution, if the equations of motion have a solution with all warp factors positive for $r>0$ and with the dilaton behaving as $g\sim r^\alpha \mathrm{e}^{-r}$ for large $r$. It turns out, that if $\mm$ is a product of spherical (or de Sitter --- in our case they result in the same equations), flat and $AdS$ factors, then a necessary and sufficient condition on the existence of an LST-like solution is the absence of $AdS$ factors (factors with negative curvature).  The curvature of positively curved factors (with a positive beta function on the worldsheet) grows towards small $r$, so that they can be weakly curved for large $r$, and shrink to zero or to finite size at $r=0$.  On the other hand, the curvature of negatively curved factors grows towards large $r$, so that the corresponding solutions do not have a smooth asymptotic large $r$ region, that can be used to define such theories.

When we only have factors of dimension $d_a$ with positive curvature, and factors of dimension $d_b$ with vanishing curvature, we can find the following asymptotic solution at $r\to \infty$:
\begin{align}
g & = g_0 r^{\sum d_a /8} \mathrm{e}^{-r}, \nonumber\\ 
c_2 &= c_3 = \sqrt{N}g^{-1/4}, \nonumber\\
\zeta_a &= \sqrt{d_a-1} \sqrt{r}, \nonumber \\
%
\zeta_b &= K_b. \label{LSTform}
\end{align}
This satisfies the equations of motion (and the constraint) in the $r\rightarrow \infty$ limit, up to terms that go as negative powers of $r$ compared to the leading terms; in particular, the constraint evaluated on this solution behaves as $\sim 1/r^2$. For every solution there is always a freedom (not just in the asymptotic region) of multiplying $g$ by a constant, since the equations are invariant under $g\to c\cdot g$; this is related to the parameter $g_0$ above.
Similarly, the string frame metric in the flat components can be any constant $K_b$ (on the worldsheet these can be just decoupled sigma models on $\mathbb{R}^n$ or $T^n$).

Whenever one of the components of spacetime has negative curvature (for instance an $AdS$ factor) there are no such solutions, and the corresponding warp factor $\zeta$ cannot remain positive in the asymptotic region. Thus, we conclude that LSTs cannot be put on spacetimes with negative curvature (in the absence of any additional background fields).

There are a few special cases where the asymptotic solutions \eqref{LSTform} are actually exact. One is the well-known flat space solution, for ${\cal M} = \mathbb{R}^d \times T^{6-d}$, when there are no curved factors. The others are new solutions that arise when the curved spaces are $S^4$ or $S^2\times S^2$, with the other two directions flat (compact or infinite).
These new exact solutions are, however, somewhat peculiar --- the dilaton vanishes at the origin $r=0$, and the string frame metric is singular there, so they do not belong to the families of smooth solutions that we mentioned above (and indeed we cannot trust them because they are singular). Nevertheless, as we will discuss below, the form of these solutions will be useful for finding and renormalizing the smooth solutions for these specific manifolds. 

A general solution will look like \eqref{LSTform} asymptotically, but will then get corrections for finite $r$. There are two classes of corrections -- power-law corrections, which appear in the non-normalizable components of the solutions, and exponential corrections, which appear in the normalizable components. 

First, in every case except for the special cases discussed above, there are power-law corrections to \eqref{LSTform} in $1/r$, and we can find them order by order in $1/r$; they take the form 
\begin{align}
g &= g_0 r^{\sum d_a/8} \mathrm{e}^{-r} \left[ 1 + \sum\limits_{i=1}^\infty \sum\limits_{j=0}^i \gamma_{i,j}\frac{(\log (r))^j}{r^i} \right], \nonumber \\
\zeta_a &= \sqrt{d_a-1} \sqrt{r} \left[ 1+ \sum\limits_{i=1}^\infty \sum\limits_{j=0}^i \alpha_{i,j}^a \frac{(\log (r))^j}{r^i} \right], \label{corrections}\\
\zeta_b &= K_b. \nonumber
\end{align}
The form of the equations implies that the $\alpha_{1,0}^a$ parameters, appearing in the leading order correction to the curved warp factors, are arbitrary, but that the rest of the solution (all the other $\alpha_{i,j}^a$'s, and the $\gamma_{i,j}$'s) can be determined in terms of these parameters.
 
So naively it seems that we have as parameters $g_0$, and one size parameter for every factor in our space, $\alpha_{1,0}^a$ or $K_b$, which we can identify with the size of that space. However, one combination of the $g_0$ and $\alpha_{1,0}^a$ parameters is redundant, since we can swallow it by shifting the coordinate $r$ (such a shift changes $g_0$, and shifts all the $\alpha_{1,0}^a$ by an equal constant). This is clear in the flat space example, where $g_0$ is not really a parameter since it can be modified by shifting the radial coordinate. The parameters $K_b$, which can be sizes of circles that the LST lives on, have an obvious physical interpretation. The interpretation of the $\alpha_{1,0}^a$ is more subtle since the size of the corresponding compact directions changes with $r$. One way to describe the physical parameters is to fix some small value of $g=g_1$ and ask for the value of the warp factors at the value of $r$ where $g(r)=g_1$.

The number of non-trivial parameters is thus the same as the number of space components, and we can think of these parameters as the sizes of the spheres that the LST is compactified on, in units of the string scale $M_s$. This is similar to the situation for the supersymmetric Maldacena-Nu\~{n}ez solutions for LST on $S^2$ \cite{MaldacenaNunez}, where the only parameter may be thought of as the size of the $S^2$ in string units \cite{NonAdS}. Note that we can always take one of these parameters to correspond to $g_0$, and that our classical solutions do not depend on $g_0$, but only the string loop corrections to them.\footnote{On the other hand, the solutions of \cite{MaldacenaNunez,MaldacenaNastase} do have a classical dependence on $g_0$, since they include also Ramond-Ramond fields, such that the equations of motion are not invariant under shifts of the dilaton.} 

In a case like $\mathbb{R}^2\times S^4$ we have a single parameter, which can be taken to be $g_0$. In this case we already mentioned that \eqref{LSTform} is an exact singular solution; we will see below that there is also a non-singular solution with the same values of the parameter. In a case like $\mathbb{R}^2\times S^2\times S^2$ we have one non-trivial parameter which has an effect in the classical limit, and which can be taken to be the difference between the $\alpha_{1,0}^a$'s of the two spheres. When this difference vanishes we have the exact singular solution \eqref{LSTform}, and we will see that for general values of this parameter (and also when it vanishes) we will have also non-singular solutions, where one of the spheres shrinks smoothly at the origin.

Since to obtain the solutions we are solving second order differential equations, we expect to have two arbitrary coefficients associated with every function (such as our warp factors). In holography generally one of these coefficients multiplies a non-normalizable mode of the corresponding field, and gives a parameter of the theory, while the other coefficient multiplies a normalizable mode, and can be thought of as a vacuum expectation value. In our configurations, the normalizable modes show up in corrections to \eqref{corrections} that behave at large $r$ as $\mathrm{e}^{-2r}$, so that they are not visible in the power series above. In each of $g$, $\zeta_a$ and $\zeta_b$, such terms can arise, and will be present in generic solutions. In particular, two solutions that have the same asymptotic parameters will differ by such exponentially decaying solutions, and will describe different configurations in the same theory. A famous example in the case of LST on $\mathbb{R}^{5,1}$ is the near-horizon near-extremal NS5-brane solutions \cite{MaldacenaStrominger}, where the dilaton and the metric in the time direction differ from \eqref{flatholo} in the large $r$ region by such asymptotically small terms.

In general it is difficult to find the explicit form of these exponentially small terms, since we do not have a closed formula even  for the leading order series \eqref{corrections}. However, in the special cases where \eqref{LSTform} is an exact solution, we can analytically find also the leading exponentially small corrections to it. Let us discuss explicitly the specific case of the theory on $\mathbb{R}\times S^1\times S^4$, where \eqref{LSTform} is an exact solution, and the general solution will look like it up to some constant shift of $r$ and up to exponentially small corrections.
In this case the most general solution (if we assume that the $\mathbb{R}$ factor describes a decoupled CFT on the worldsheet) can be written as
\begin{align} \label{AsymptoticWarpFactors}
\begin{split}
\zeta_{S^4}(r) &= \sqrt{3} \sqrt{2A + r}\left(1+ \mathrm{e}^{-2r}f_0(r)+\mathcal{O}( \mathrm{e}^{-4r})\right), \\
\zeta_{\mathbb{R}}(r) &= K_1, \\
\zeta_{S^1}(r) &= K_s\left(1+ \mathrm{e}^{-2r}f_s(r)+\mathcal{O}(\mathrm{e}^{-4r})\right), \\
g(r) &= g_0 \sqrt{3} \sqrt{2A + r}\;\mathrm{e}^{-r}\left(1+ \mathrm{e}^{-2r}f_g(r)+\mathcal{O}(\mathrm{e}^{-4r})\right).
\end{split}
\end{align}
	The leading correction functions $f_0$, $f_s$ and $f_g$ can be found exactly, up to two unknown constants $a_0$ and $a_1$, and are given by
	\begin{align}
		f_0(r) &= a_0 \frac{2A+r+1}{2A+r}, \\
		f_s(r) &= - a_1 \mathrm{e}^{2r+4A} \mathrm{Ei}(-4A-2r), \\
		f_g(r) &= f_s(r) + a_0 \frac{4A+2r+1}{2A+r}.
	\end{align}
	Here $\mathrm{Ei}(x)$ is the special function defined as
	\begin{equation}
		\mathrm{Ei}(x) = -\int\limits_{-x}^\infty \frac{\mathrm{e}^{-t} \mathrm{d} t}{t},
	\end{equation}
	where the principal value of the integral is taken; in the $-x\gg 1$ limit $\mathrm{Ei}(x) \sim -\mathrm{e}^{-x}/x$, so for large $r$, $f_s(r) \approx 1/2r + \mathcal{O}(1/r^2)$. 
	We interpret $a_0$ and $a_1$ as the coefficients of the normalizable modes associated with the sizes of the $S^4$ and the $S^1$, respectively. We expect that their values should determine also all the higher order exponential corrections, which we do not compute explicitly.
	As we will discuss below, knowing the solutions to order $\mathrm{e}^{-2r}$ will allow us to compute the free energy for these solutions.

There is one case where we were able to find a smooth exact solution, for which $\mathcal{M} = \mathbb{R} \times S^1 \times S^4$ (so it is of the general form \eqref{AsymptoticWarpFactors}). Let us consider the case when the $S^4$ sphere shrinks and the $S^1$ doesn't (this is essentially equivalent to $\mathbb{R}\times \mathbb{R}\times S^4$). In such a case, $\zeta_{S^1}(r)$ is constant, and the equations of motion simplify. After some manipulations, one can find the smooth solution
\begin{align} \label{exacts1}
\begin{split}
g(r) &= g_0 \frac{\mathrm{e}}{2}\frac{\sqrt{3} \sqrt{r \, \mathrm{coth} (r) - 1}}{\mathrm{sinh} (r)}, \\
\zeta_{S^4}(r) &= \sqrt{3} \sqrt{r \, \mathrm{coth} (r) - 1}, \\
\zeta_{S^1}(r) &= \frac{\beta}{2 \pi \sqrt{N}},
\end{split}
\end{align}
where $\beta$ is the circumference of the $S^1$ in string units.
As usual, $g_0$ is an arbitrary coefficient, whose normalization we choose here in such a way that as $r$ goes to infinity, one has
\begin{equation} \label{gzetasingular}
g(\zeta_{S^4}(r)) \approx g_0 \mathrm{\zeta}_{S^4} (r) \mathrm{e}^{-\zeta_{S^4} (r)^2/3}.
\end{equation}

The theory on $\mathbb{R}^2\times S^4$ thus has a single parameter $g_0$, which has no classical effect, and two solutions, describing different states in the theory -- the smooth solution \eqref{exacts1} and the singular one \eqref{LSTform}. It would be interesting to understand if the singularity in \eqref{LSTform} can be resolved in classical string theory (the solution is arbitrarily weakly coupled), and if either or both solutions are unstable (this can be checked by analyzing small fluctuations around the solution, and seeing if there are tachyons). There is a limit of LST on $S^4$, when the $S^4$ is much larger than ($\sqrt{N}$ times) the string scale, where the low-energy theory should describe (for the ${\cal N}=(1,1)$ LST on $S^4$) the 6d $\mathrm{SU}(N)$ supersymmetric Yang-Mills theory on $S^4$, which reduces at low energies to a 2d Yang-Mills theory. As in \cite{MaldacenaNunez,MaldacenaNastase}, in this limit $g_0$ becomes large \cite{NonAdS} such that the classical solutions are no longer reliable.

\subsection{Series solution}


As we mentioned, to find the numerical solutions, it is convenient to start from analytic solutions as a power series near $r=0$, and then to continue them into numerical solutions for larger values of $r$. We look for solutions where space ends smoothly at $r=0$, which means that some compact $S^n$ factor smoothly shrinks there, with a local metric proportional to $(\mathrm{d} r^2 + r^2 \mathrm{d} \Omega_n^2)$. Here $\mathrm{d} \Omega^2_n$ is the metric on the unit $n$-sphere, and we can have $n=1$ if a flat circle factor shrinks to zero size, or $n>1$ if a curved sphere shrinks. In any case there is one special warp factor $\zeta_0$ (which can be one of the $\zeta_a$'s, or one of the $\zeta_b$'s corresponding to an $S^1$) for which we need to take $\zeta_0(0)=0$, $\zeta_0'(0)=1$. All other warp factors go to a constant size there, and to get a solution we need to take $\zeta_k(0)=\rho_k$, $\zeta_k'(0)=0$. Similarly, the dilaton can go to some arbitrary constant $g(0)$, and without loss of generality we can always choose $g(0)=1$ (since we can always multiply $g(r)$ by an arbitrary constant and still have a solution). Solving the EOM then requires $g'(0)=0$. This gives us a full list of initial conditions for our differential equations, which satisfy the constraint, and it is easy to check that these are the most general possible values for which space ends with a finite curvature. Thus, our solutions are parameterized by the sizes $\rho_k$ of all the non-vanishing warp factors at $r=0$, and by $g(0)$ which we can add in at the end.

Note that once we chose which warp factor vanishes (equivalent to a choice of topology), the number of parameters of our solutions near the origin is precisely the same as the number of asymptotic parameters. So naively we may expect to find one smooth solution for each value of these asymptotic parameters. However, as we will discuss in sections \ref{sec:hagedorn} and \ref{oscs}, the correspondence between the parameters at infinity and at the origin is not one-to-one. For some asymptotic parameters there is no solution with a specific topology, while for others there is more than one solution for the same asymptotic parameters and the same topology.

	
To find the series solution we plug the series form
	\begin{equation}
		\zeta_k(r) = \zeta_k(0) + \zeta'_k(0) r + \sum_{n=2}^{\infty} \frac{z_{n,k} r^n}{n!}, \quad g(r) = g(0) + g'(0)r + \sum_{n=2}^{\infty} \frac{\nu_n r^n}{n!}
	\end{equation}
	into the equations of motion, and expand the equations of motion and the constraint around $r=0$. This turns differential equations into a set of linear equations that we can solve. We solve these up to some power $r^n$. These equations become more and more cumbersome with the growth of $n$, so at a certain point we cease solving them and use the resulting expansion to get $\zeta_k(r_0)$, $\zeta'_k(r_0)$, $g(r_0)$ and $g'(r_0)$, at some finite value of $r_0$ for which the $r_0^n$ corrections are very small. We then solve the equations numerically starting from these new initial values.

\subsection{Numerical solution} \label{numsol}


We obtain the numerical solution using the NDSolve procedure of Mathematica, starting almost at the origin, e.g. at $r_0=10^{-6}$ (the initial conditions are obtained using the series expansion near zero, as explained in the previous section). Increasing the values of the AccuracyGoal and PrecisionGoal parameters is relevant for increasing computational precision; however, it is much more important for the computation of the free energy which we will discuss below. 

Let us present for illustration two plots of the warp factors. The warp factors for $S^3 \times S^3$ (the initial conditions are $\zeta_0(0) = 1$, $\zeta_1(0) = 0$) are shown in Figure \ref{fig:S3S3nonsingular}:

\begin{figure}[H]
\includegraphics[width=9cm]{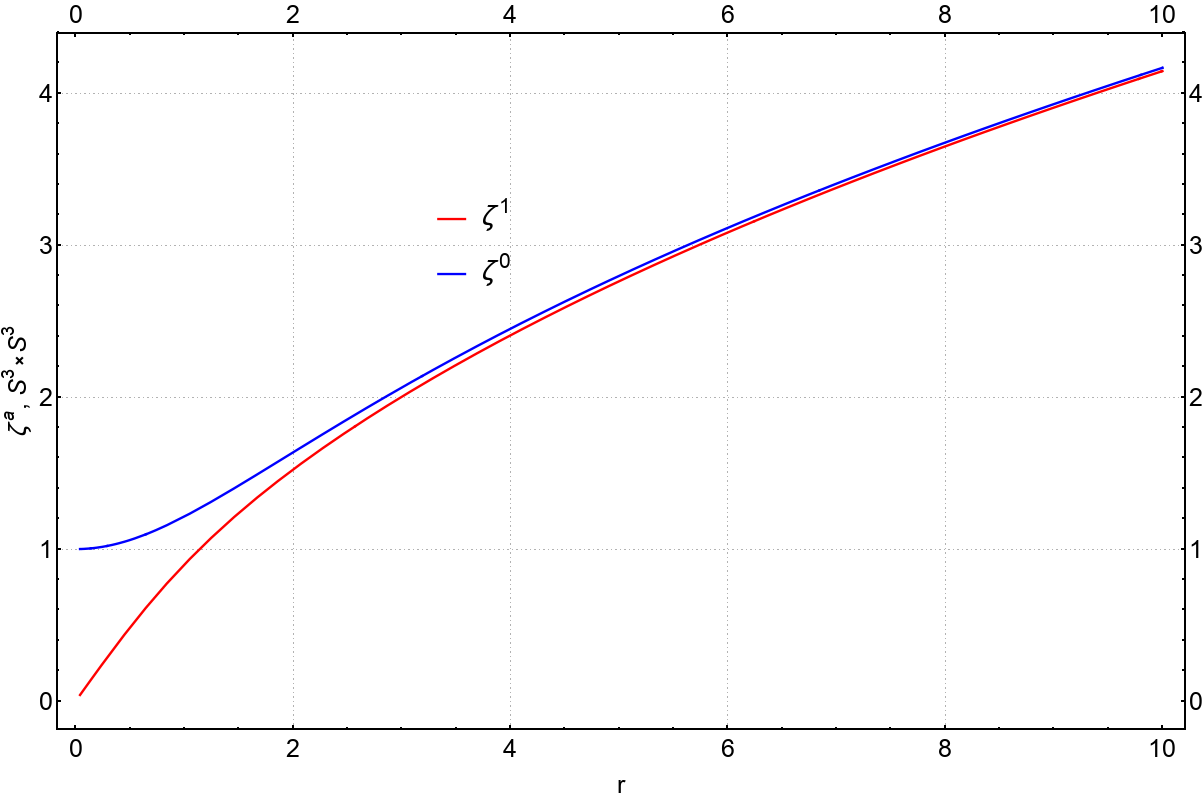} \centering \caption{The warp factors for $\mathcal{M} = S^3 \times S^3$.} \label{fig:S3S3nonsingular}
\end{figure}

The warp factor $\zeta_4$ corresponding to $S^4$ in LST on $\mathbb{R}^2\times S^4$ is shown in Figure \ref{fig:R2S4nonsingular}, where we have also plotted $\sqrt{3r}$ for comparison, as for large $r$, $\zeta_4(r) \approx \sqrt{3r}$ (and $\sqrt{3r}$ is an exact, but singular solution):

\begin{figure}[H]
\includegraphics[width=9cm]{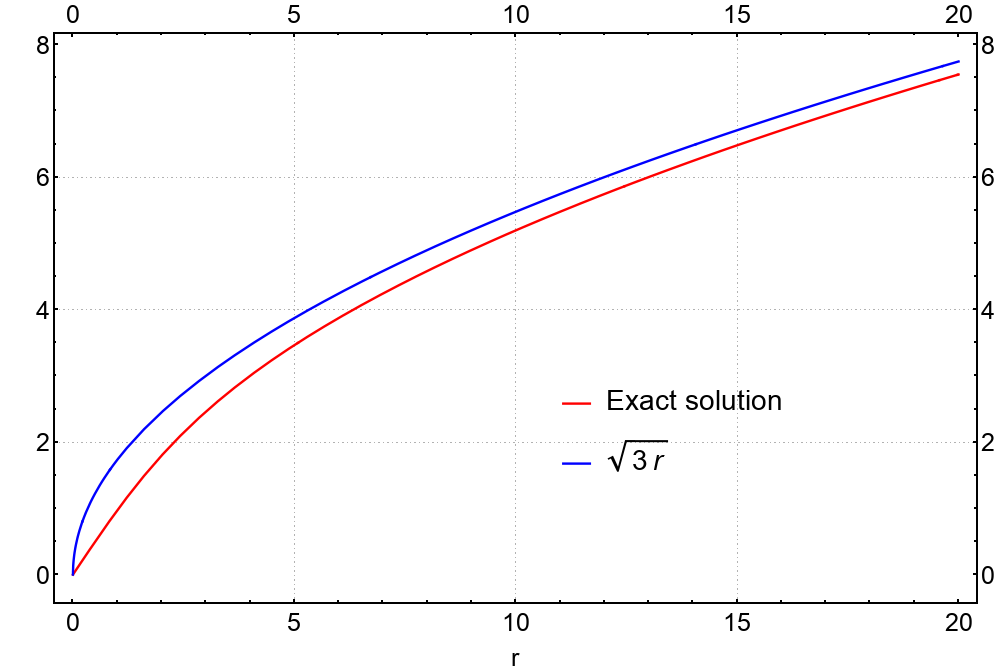} \centering \caption{The $S^4$ warp factor for $\mathcal{M} = \mathbb{R}^2 \times S^4$.} \label{fig:R2S4nonsingular}
\end{figure}

NDSolve usually breaks down at some radial position $r\sim 1000$, when the dilaton value becomes very small, so one cannot get a numerical solution all the way to infinity. However, at this value of $r$ the higher-order corrections to the asymptotic solutions are negligible (we considered terms up to $1/r^4$ order, getting $10^{-12}$ precision), so we can sew together the numerical and the asymptotic solutions. Sewing is done by minimizing 
\begin{align}
\begin{split}
	\sum_k &(c_{asymp, k}(r_1) - c_{num, k}(r_1))^2 + \sum_k (c_{asymp, k}(r_2) - c_{num, k}(r_2))^2 +  \\
	+ &\left( \frac{g_{asymp}(r_1)}{g_{num}(r_1)} - 1 \right)^2 + \left( \frac{g_{asymp}(r_2)}{g_{num}(r_2)} - 1 \right)^2
\end{split}
\end{align} 
with respect to the parameters of the asymptotic solution, $g_0$ and $\alpha_{1,0}^a$. We take two different points $r_1$, $r_2$ that are far from zero (e.g. 990 and 900), in order to ensure that the solutions indeed coincide and do not merely intersect. 

It is important to verify that for a large range of values of $r$ we have agreement between the numerical and asymptotic solutions. For example, in the $S^3 \times S^3$ case, where we got the asymptotic solution with the corrections up to order $\mathcal{O} \left( 1/r^3 \right)$, we plot the relative error between the two in Figure \ref{fig:Sewingprecision}:

\begin{figure}[H]
\includegraphics[width=11cm]{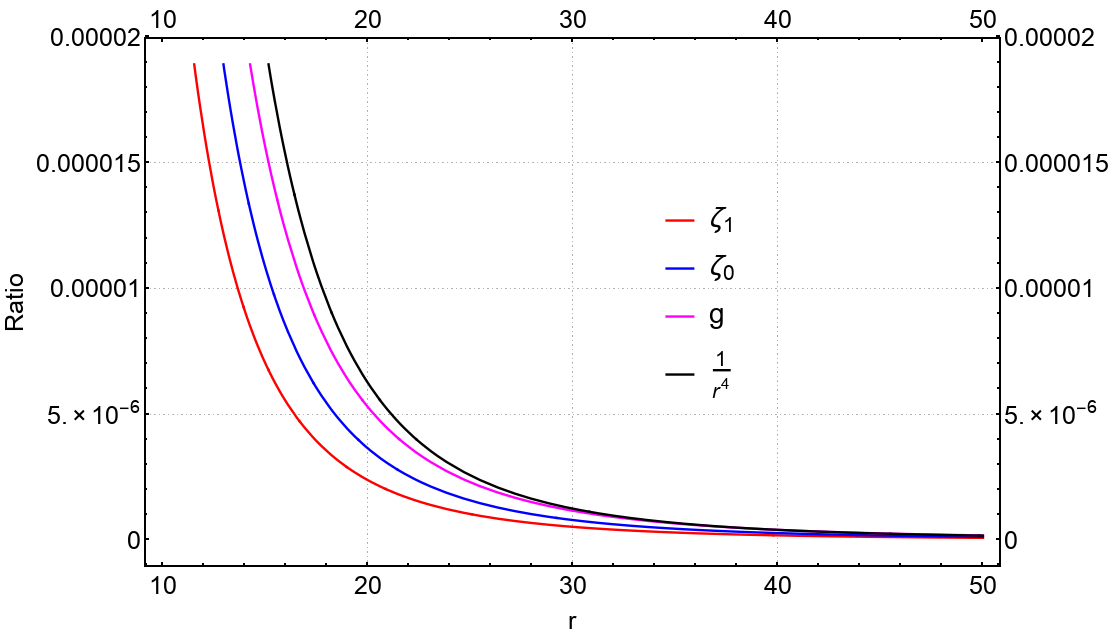} \centering \caption{Relative sewing precision for $\mathcal{M} = S^3 \times S^3$.} \label{fig:Sewingprecision}
\end{figure}

In this figure $\zeta_1$, $\zeta_0$ and $g$ correspond to the ratio 
\begin{equation}
\frac{f_{num} - f_{asymp}}{f_{asymp}},
\end{equation}
where $f$ is correspondingly $\zeta_1$, $\zeta_0$ (the warp factors of the two $S^3$ factors) or $g$, and the $num, asymp$ subscripts correspond to the numerical and asymptotic solutions. One can see that the asymptotic and numerical solutions are indeed very close to each other starting from some value of $r$, so we can safely sew them together; as $g$ is very small for large $r$, we look for $\frac{g_{asymp}}{g_{num}}$ to be close to one.

When we have more than one compact manifold, we solve the equations numerically for various values of the non-vanishing radii at $r=0$, and of course the $\alpha_{1,0}^a$ coefficients that we find depend on these. These asymptotic parameters are defined up to the freedom of shifting $r$, and one combination that is invariant under this is the difference between the $\alpha^a_{1,0}$, so it is meaningful to ask how this depends on the initial sizes at $r=0$. The result for the difference $\alpha^0_{1,0} - \alpha^1_{1,0}$ in the case of $\mathcal{M} = S^3 \times S^3$, as a function of the initial size $\rho$ for one of the $S^3$'s, is shown in Figure \ref{fig:S3S3largerho}:

\begin{figure}[H]
\includegraphics[width=9cm]{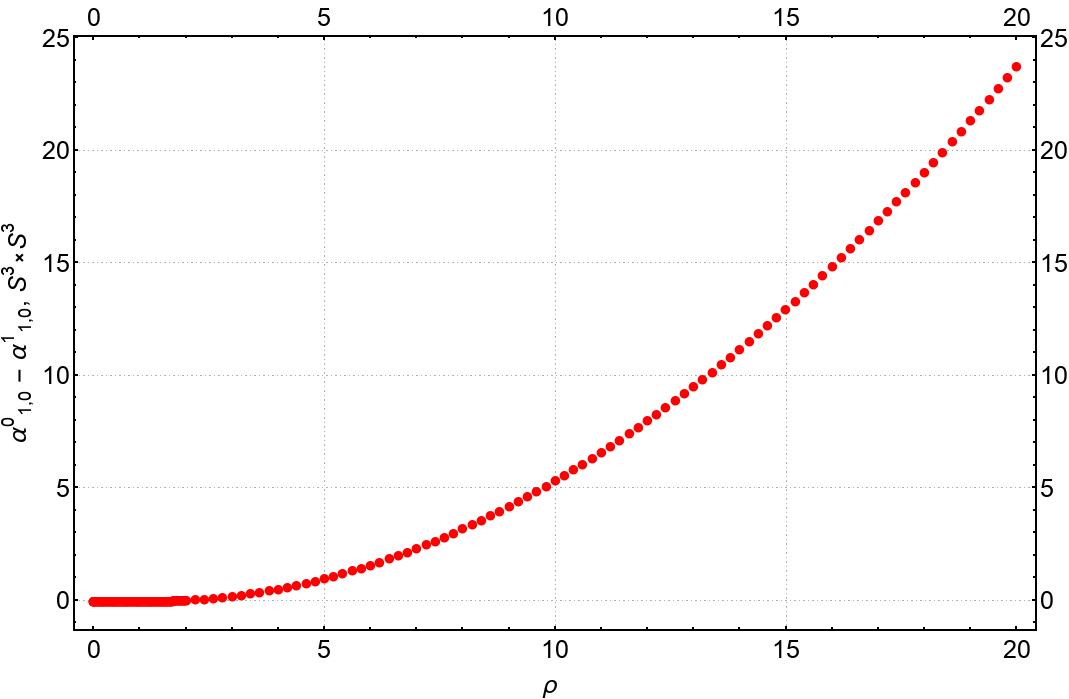} \centering \caption{The difference $\alpha_{1,0}^0 - \alpha_{1,0}^1$ for $\mathcal{M} = S^3 \times S^3$.} \label{fig:S3S3largerho}
\end{figure}

\begin{figure}[H]
\includegraphics[width=9cm]{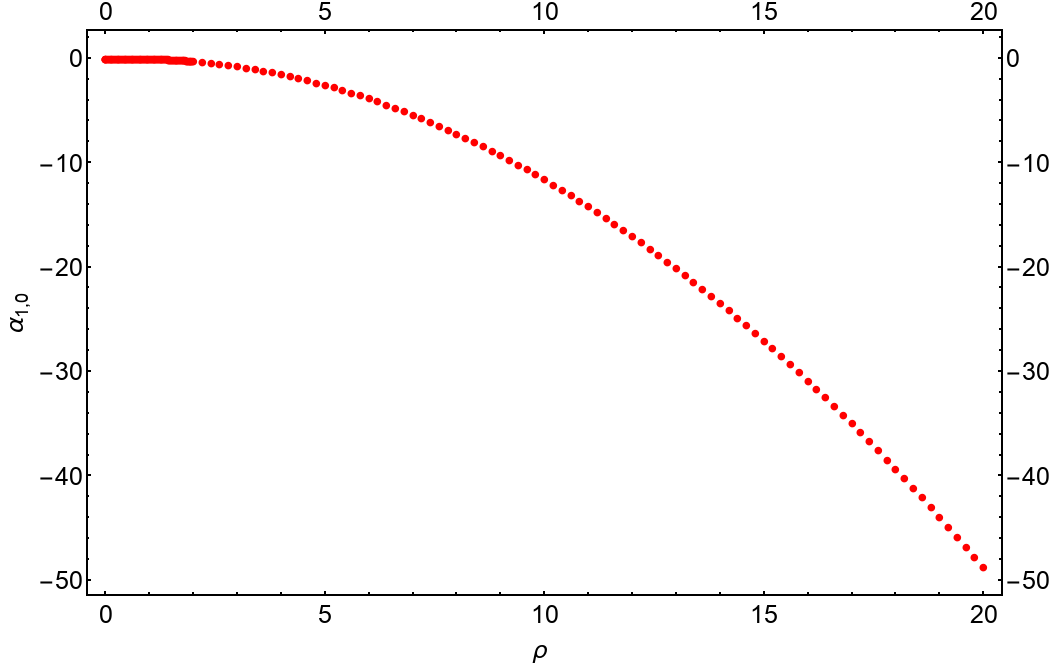} \centering \caption{The difference $\alpha_{1,0}^4 - \alpha_{1,0}^2$ for $\mathcal{M} = S^2 \times S^4$, where $S^4$ shrinks.} \label{fig:S2S4largerhoS4shrinks}
\end{figure}

We have also studied how the $\alpha_{1,0}$ coefficients depend on $\rho$ for the $S^2 \times S^4$ case (for two different options of either $S^2$ or $S^4$ shrinking; the upper index of $\alpha_{1,0}$ corresponds to the dimension of the sphere). The results appear in Figures \ref{fig:S2S4largerhoS2shrinks} and  \ref{fig:S2S4largerhoS4shrinks}:

\begin{figure}[H]
\includegraphics[width=9cm]{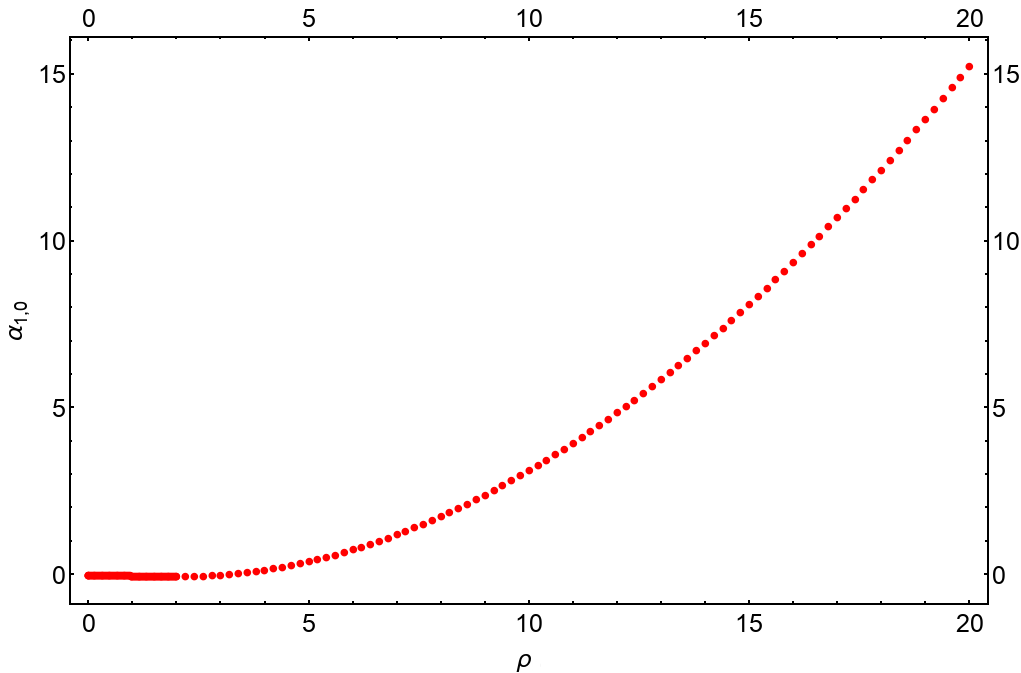} \centering \caption{The difference $\alpha_{1,0}^4 - \alpha_{1,0}^2$ for $\mathcal{M} = S^2 \times S^4$, where $S^2$ shrinks.} \label{fig:S2S4largerhoS2shrinks}
\end{figure}


At first glance these differences look monotonic, but this is merely an artifact of the plots. In fact, for all the cases shown in Figures \ref{fig:S3S3largerho}-\ref{fig:S2S4largerhoS4shrinks} the asymptotic parameter $\alpha^a_{1,0} - \alpha^b_{1,0}$ exhibits oscillating behavior in the small $\rho$ region, as we will discuss in section \ref{oscs}. When one sphere is asymptotically much larger than the other, there is only a solution in which the smaller sphere shrinks smoothly, but when the radii are comparable, more than one solution may exist.


\section{The free energy and thermodynamics}\label{renorm}

One interesting property of our solutions is their Euclidean classical action, which is the same as the free energy divided by the temperature when the background has an $S^1$ factor. When we have several solutions with the same asymptotic behavior, the one with the lowest Euclidean action will dominate the path integral in the classical limit, and we can have phase transitions between different dominant configurations as we change the parameters.

In our calculation of the action, we follow Cotrone et al. \cite{CotronePonsTalavera}, who have found that the free energy of LST in flat space vanishes. 
As usual in holographic backgrounds, we will need to put a cutoff on the radial direction, and to add some local counterterms  in order to make the action well-defined. In order to analyze this it is enough to look at the asymptotic solutions \eqref{LSTform}.
The action contains the volume part \eqref{baction} and the surface Gibbons-Hawking term
\begin{equation}
	I_{GH} = -\frac{1}{\varkappa^2} \int \mathrm{d}^{9} x \sqrt{h} K = - \frac{1}{\varkappa^2} \int \mathrm{d}^{9} x \sqrt{h} \frac{1}{\sqrt{G}} \partial_{\mu} \left( \sqrt{G} n^{\mu} \right), \label{saction}
\end{equation}
where $n^\mu = \delta_{r}^{\mu}/\sqrt{G_{rr}}$ is the boundary outward normal unit vector, and $h$ is the absolute value of the determinant of the metric $h_{ab}$ induced on the boundary. All in all, the action with some cutoff $L$ is given by\footnote{Here and in what follows, we use the units of length in which $2 \varkappa^2 = 1$, unless otherwise stated.}
\begin{equation}
	I(L) = -2 \left. \int \mathrm{d}^{9} x \sqrt{h} \frac{1}{\sqrt{G}} \partial_{\mu} \left( \sqrt{G} n^{\mu} \right) \right\vert_{r=L} - \int^L \mathrm{d}^{10} x \sqrt{G} \left( R - 2N^2 \frac{1}{c_3^6 g} - \frac{1}{2 }\frac{g^{\prime 2}}{g^2 c_2^2} \right). \label{faction}
\end{equation}
When we substitute the asymptotic solution \eqref{LSTform} into \eqref{faction}, we get a mixture of exponents, polynomial factors and special functions. If we expand around $r=\infty$, the leading terms take the form (after performing the integration over the radial coordinate)\footnote{We disregard the volume of the flat coordinates, and some constant factors for each sphere: all the warp factors and the dilaton depend only on the radial coordinate, so all terms in the action and the counterterms have a common factor $\int_{\partial M} \mathrm{d}^9 x$, which we drop.}
\begin{equation}
	I[c_i^{asymp}] = \left. \mathrm{e}^{2r} r^{\sum d_a/4 - 1} \left(w_1 r +w_0  + \sum\limits_{i=1}^\infty q_i r^{-i} \right)\right\vert_{r=r_{min}}^{r=L}, \label{actionexp}
\end{equation}
where $w_i, q_i$ are some numerical coefficients. In general all $q_i \neq 0$, and they all give diverging contributions to the action (as $L\to \infty$) which must be canceled by counterterms. Thus, depending on the precise form of these coefficients, an infinite number of counterterms may be needed, and we need to know all power-law corrections to \eqref{LSTform} in order to compute them. Since we do not know them, we cannot perform the holographic renormalization procedure in general.

However, for the few cases where we have exact solutions, all $q_i$ vanish. For instance, for the $\mathbb{R} \times S^1 \times S^4$ case \eqref{AsymptoticWarpFactors}, the action takes the form
\begin{equation}
	I[c_i^{asymp}] = \left. \left( - \mathrm{e}^{2r} \left[\frac{72 K_s}{g_0^2}r + \frac{36K_s(1+4A)}{g_0^2} \right] + w_f + \mathcal{O} \left( \mathrm{e}^{-2 r} \right) \right) \right\vert_{r=r_{min}}^{r=L} \label{actioncando}
\end{equation}
for some constant $w_f$. The counterterms are surface terms and so should be constructed from the metric induced on the nine-dimensional boundary, and from the dilaton there. It turns out that for all the cases when it is possible to explicitly compute the divergent terms of the action, one only needs the two following counterterms to regularize it:
\begin{align}
	I_1(L) &= -\int\limits_{\partial \mathcal{M}} \mathrm{d}^9 x \sqrt{h}g^{1/4}, \\
	I_2(L) &= -\int\limits_{\partial \mathcal{M}} \mathrm{d}^9 x \sqrt{h} R_h g^{-1/4}. \label{cterms}
\end{align}
Here $R_h$ is the intrinsic curvature on the boundary at $r=L$; $I_1$ is similar to $\mathcal{I}_{ctgravity}$ in \cite{CotronePonsTalavera}. For all cases where \eqref{LSTform} is an exact solution, we have found that the action after adding the counterterms vanishes, similarly to the $\mathbb{R}^6$ case studied by Cotrone et al.

We can now compute the action for any solution that has the same asymptotics as our exact solutions. For instance, for
$\mathbb{R} \times S^1 \times S^4$, the renormalized action can be written as
\begin{align}
	I \approx & \left.I[c_i^{asymp}]\right\vert_{r=\Theta}^L + \kappa_1 I_1(L) + \kappa_2 I_2(L) - \nonumber \\
	&-\left. \int \mathrm{d}^{10} x \sqrt{G} \left( R - 2N^2 \frac{1}{c_{3}^6 g} - \frac{1}{2} \frac{g^{\prime 2}}{g^2 c_{2}^2} \right)\right\vert_{r=0}^\Theta. \label{freeenergy}
\end{align}
Here $\Theta$ is some large value of $r$, for which the exponentially small terms in \eqref{actioncando} can be neglected, $L$ is the cutoff on the radial coordinate, $\kappa_1$ and $\kappa_2$ are the counterterm coefficients, and the last term is the action evaluated on the numerical solution of the equations of motion in the region where we cannot neglect the exponential corrections to the solutions. Substituting \eqref{AsymptoticWarpFactors} into \eqref{freeenergy}, we find that to cancel the dependence on $L$, one has to choose $\kappa_1 = -1$ and $\kappa_2 = -1/2$, and then \eqref{freeenergy} can be rewritten as 
\begin{align}
\begin{split}
	- \frac{9 \left[(8A -4)a_0 + a_1 \right] K_s}{2g_0^2} - \left. \int \mathrm{d}^{10} x \sqrt{G} \left( R - 2N^2 \frac{1}{c_{3}^6 g} - \frac{1}{2} \frac{g^{\prime 2}}{g^2 c_{2}^2} \right)\right\vert_{r=0}^\Theta + \mathcal{O} \left( \mathrm{e^{-2 \Theta}} \right).
\end{split}
\end{align}
All the divergences have been canceled, and to compute the free energy we just need to compute the numerical solution, sew it with the asymptotic solution, and plug the results into the equation above. 

The procedure above sometimes leads to large numerical errors. These can be decreased by defining a function
\begin{equation}
	L_{v} = \frac{\mathrm{d}}{\mathrm{d} L} \left(I_{GH}(L) - I_1(L) - \frac{1}{2}I_2(L) \right).
\end{equation}
Then
\begin{equation}
	\int \mathrm{d}^{10} x L_v = \left.\left(I_{GH} - I_1 - \frac{1}{2}I_2 \right)\right|_{r=L} - \left.\left(I_{GH} - I_1 - \frac{1}{2}I_2 \right)\right|_{r=0},
\end{equation}
where the first term is the surface term and the counterterms we add to the bulk action, and the second term is a constant that doesn't depend on the cutoff value. This means that to compute the action, instead of \eqref{freeenergy} we can consider the expression 
\begin{equation}
	- \int \mathrm{d}^{10} x \left[\sqrt{G} \left( R - 2N^2 \frac{1}{c_3^6 g} - \frac{1}{2} \frac{g^{\prime 2}}{g^2 c_2^2} \right) + L_v\right] - \left.\left(I_{GH} - I_1 - \frac{1}{2}I_2 \right)\right|_{r=0}, \label{betterfree}
\end{equation}
in which the expression inside the integral by construction won't diverge exponentially. This helps us to reduce the numerical error, and to obtain the answer by evaluating the integral above numerically from zero to some cutoff value $L$; we present some results in the next section.




\subsection{Corrections to the Hagedorn spectrum} \label{sec:hagedorn}

In flat space, at leading order in the string coupling the thermodynamics of LST leads to an exact Hagedorn spectrum \cite{BerkoozRozali, HarmarkObers}\footnote{See \cite{Rangamani,Buchel:2001dg,NarayanRangamani,DeBoerRozali,AharonyGiveonKutasov,ParnachevStarinets,HarmarkObers2,ParnachevSahakyan,HarmarkNiarchosObers,BarbonFuertesRabinovici,BarbonRabinovici,LorenteEspinTalavera,BertoldiHoyosBadajoz,LorenteEspin,Sugawara} for additional works on the thermodynamics of little string theories.}
\begin{equation}
S = \beta_H E,
\end{equation}
and at one-loop order this is corrected as  \cite{KutasovSahakyan}
\begin{equation} \label{hagcorr}
S = \beta_H E + \alpha \log \left( \beta_H E\right) + o \left( \log \left( \beta_H E\right) \right),
\end{equation}
where $\beta_H \equiv 2 \pi \sqrt{N \alpha'}$, and $\alpha$ is a negative coefficient with linear dependence on the volume of the five spatial coordinates (assuming large volume compared to the string scale).

In curved space the situation is different, and it was found already in \cite{Buchel:2001dg,Gubser:2001eg,BuchelInstability} that there are corrections to the Hagedorn spectrum of LST on $S^2$ already at the classical level. We expect such corrections to appear on general curved spaces; here we analyze in detail the case of LST on $S^4$ at finite temperature, for which we can numerically compute the renormalized free energy as described above.\footnote{In this paper we discuss only the classical contribution to the thermodynamics, related to the classical action and, in black hole solutions, to the horizon area of the black hole. There are also important contributions to the thermodynamics at one-loop order coming from the thermal fluctuations of the various fields in the holographic background, and because the radial direction is infinite, these could significantly modify the analysis in the canonical ensemble \cite{BarbonFuertesRabinovici}. It would be interesting to analyze these contributions in our curved space solutions.} This corresponds to the Euclidean theory on  $\mathcal{M} = \mathbb{R} \times S^1 \times S^4$, where the asymptotic circumference of the circle is identified with the inverse temperature $\beta=1/T$, and the Euclidean action is identified with $\beta$ times the free energy. We expect two types of solutions; one where the $S^4$ shrinks and the $S^1$ has constant radius, corresponding to a thermal gas of the original particles of LST on $S^4$, and a black hole solution where the $S^1$ shrinks to zero size at the horizon. For the black hole solutions we can analyze the thermodynamics using
\begin{equation}
\mathrm{d} E = T \mathrm{d} S,
\end{equation}
where $S = \mathrm{A}/4G_N$ is the entropy of the black hole with the horizon area $\mathrm{A}$. In this subsection, we work in units where $\alpha' = 1$, and $16 \pi G_N = (2 \pi)^7 g_s^2 \alpha'^4 = (2 \pi)^7 g(0)^2$. As usual in the linear dilaton setup, the string coupling in the relation above is taken to be the value of the coupling at the horizon (the minimal radial coordinate), $g_s = g(0)$.

The field equations we solve here are\footnote{The warp factor corresponding to the $\mathbb{R}$ factor is constant and doesn't show up in the equations of motion.}
\begin{align} \label{EoMs}
\begin{split}
&\left(\frac{g'}{g} \frac{\zeta_0^4 \zeta_1}{g^2} \right)' + 2 \frac{\zeta_0^4 \zeta_1}{g^2} = 0, \\
&\left(\frac{\zeta_0'}{\zeta_0} \frac{\zeta_0^4 \zeta_1}{g^2} \right)' - 3 \frac{\zeta_0^4 \zeta_1}{g^2} \frac{1}{\zeta_0^2} = 0, \\
&\left(\frac{\zeta_1'}{\zeta_1} \frac{\zeta_0^4 \zeta_1}{g^2} \right)' = 0,
\end{split}
\end{align}
where $\zeta_0$ and $\zeta_1$ are the warp factors of the four-sphere and the circle, respectively. The constraint is
\begin{equation} \label{constr}
- 3 \frac{1}{\zeta_0^2} + 3 \frac{\zeta_0'^2}{\zeta_0^2} + 2 \frac{\zeta_0' \zeta_1'}{\zeta_0 \zeta_1} - 4 \frac{\zeta_0' g'}{\zeta_0 g} - \frac{\zeta_1' g'}{\zeta_1 g} + \frac{g'^2}{g^2} = 1.
\end{equation}

For the black hole solutions the boundary conditions at the horizon $r=0$ are labeled by the size of the $S^4$ at the horizon, which we denote by $\rho$, and by the string coupling there:
\begin{align}
\begin{split}
g(0) &= \rho f(\rho), \quad g'(0) = 0, \\
\zeta_0(0) &= \rho, \quad \zeta_0'(0) = 0, \\
\zeta_1(0) &= 0, \quad \zeta_1'(0) = 1,
\end{split}
\end{align}
where $f(\rho)$ is a function which we will specify momentarily. These are consistent with the constraint \eqref{constr}.

We expect our solution to connect smoothly with the asymptotic solution (\ref{AsymptoticWarpFactors}) at infinity, so we propose the following ansatz:
\begin{align}
\begin{split}
g(r) &= \sqrt{2A(\rho) + r} \cdot \mathrm{G}(r,\rho), \label{hagedorn}\\
\zeta_0(r) &= \sqrt{2A(\rho) + r} \cdot \mathrm{F}(r,\rho), \\
\zeta_1(r) &= \mathrm{H}(r,\rho),
\end{split}
\end{align}
where the functions $\mathrm{G}$, $\mathrm{F}$, and $\mathrm{H}$ are everywhere bounded from above.

We didn't manage to solve the equations of motion exactly, but for large values of the radius $\rho^2$ at the horizon, it is possible to construct a solution as an expansion in $1/\rho^2$ (as in \cite{BuchelInstability}). Since we expect the solution for $\mathrm{G}$, $\mathrm{F}$, and $\mathrm{H}$ to be bounded, the expansion in $1/\rho^2$ must be uniform. We compute the first three orders in $1/\rho^2$ expansion, and find that it is indeed the case. We also verify our claims numerically.

In order to use our ansatz, we must find the dependence of $A$ on $\rho$. This may be done by properly sewing the numerical solution and the asymptotic solution (\refeq{AsymptoticWarpFactors}). We used another way to find it, namely we parameterized $A(\rho)$ as
\begin{equation}
A(\rho) = \frac{\rho^2}{6} \left[1 + \frac{A_0}{\rho^2} + \frac{A_2}{\rho^4} + \frac{A_4}{\rho^6} + \mathcal{O} \left( \frac{1}{\rho^8} \right) \right],
\end{equation}
and required that the dependence on $r$ is consistent with (\refeq{AsymptoticWarpFactors}). We found the solution up to the third order in $1/\rho^2$ expansion, which allowed us to determine $A_0$ and $A_2$ to be
\begin{equation} \label{a02}
A_0 = - 3 \log (2), \qquad A_2 = \frac{3}{8} \left( 12 \log (2) - \pi^2 \right).
\end{equation}
This matches nicely with the $A(\rho)$ data obtained from sewing together the asymptotic solution \eqref{AsymptoticWarpFactors} to the numerical ones, see Figure \ref{fig:Arhocomparison}:

\begin{figure}[H]
\includegraphics[width=9cm]{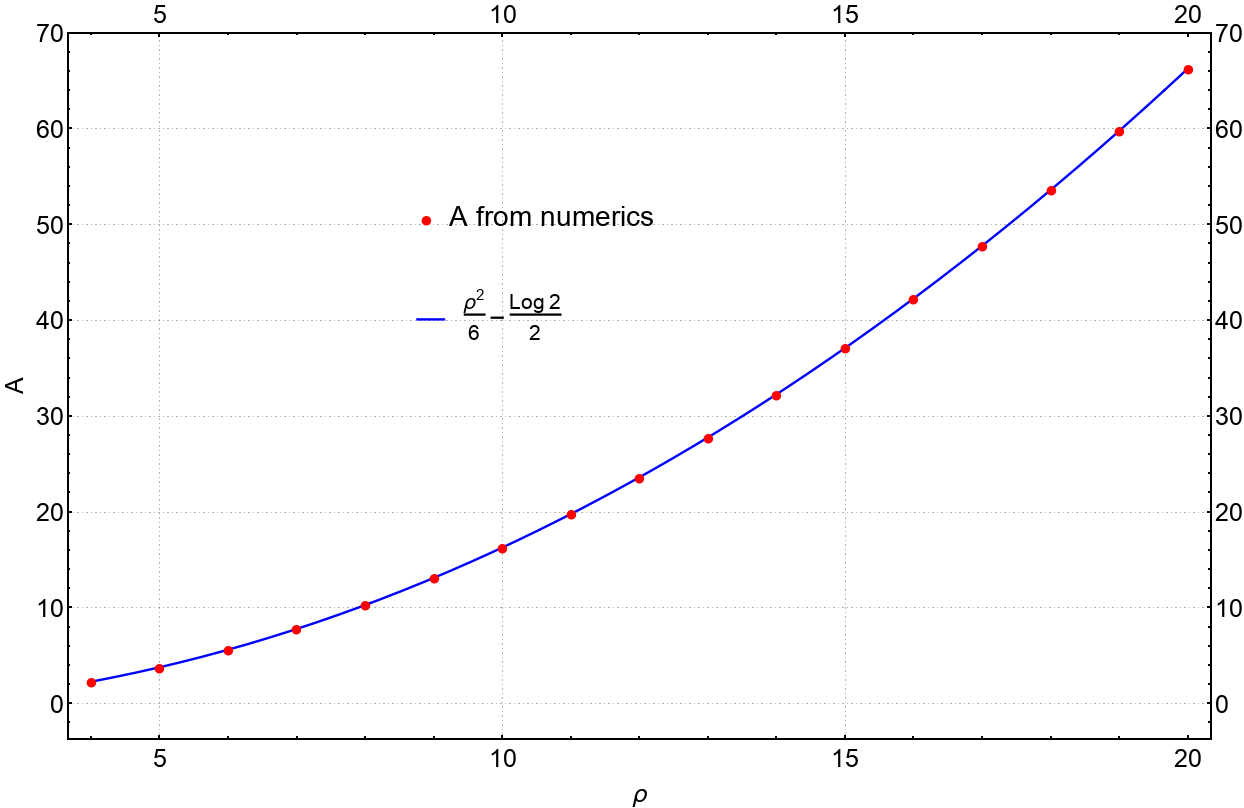} \centering \caption{The dependence $A (\rho)$ for large $\rho$, found numerically and compared to \eqref{a02}.} \label{fig:Arhocomparison}
\end{figure}

Next, substituting the functions $\mathrm{G}$, $\mathrm{F}$, and $\mathrm{H}$ in the form of a power series in $1/\rho^2$, we find a solution of the form
\begin{align}
\begin{split}
g(r) &= \sqrt{3} \sqrt{2A(\rho) + r} f(\rho ) \frac{1}{\mathrm{cosh} (r)} \left[ 1 + \frac{G_2 (r)}{\rho^2} + \frac{G_4 (r)}{\rho^4} + \mathcal{O} \left( \frac{1}{\rho^6} \right) \right], \\
\zeta_0(r) &= \sqrt{3} \sqrt{2A(\rho) + r} \left[ 1 + \frac{F_2 (r)}{\rho^2} + \frac{F_4 (r)}{\rho^4} + \mathcal{O} \left( \frac{1}{\rho^6} \right) \right], \\
\zeta_1(r) &= \mathrm{tanh} (r) \left[ 1 + \frac{H_2 (r)}{\rho^2} + \frac{H_4 (r)}{\rho^4} + \mathcal{O} \left( \frac{1}{\rho^6} \right) \right]. \label{hyperform}
\end{split}
\end{align}
Here
\begin{align}
\begin{split}
G_2 (r) &= \frac{3}{2} \left\{ 2 \log \left( \mathrm{cosh} (r) \right) - r \left[ 1 + \mathrm{tanh} (r) \right] + \log (2) \right\}, \\
F_2 (r) &= \frac{3}{2} \left\{ \log \left( \mathrm{cosh} (r) \right) - r + \log (2) \right\}, \\
H_2(r) &= \frac{3}{2} \left\{ \frac{2r}{\mathrm{sinh}(2r)} - 1 \right\},
\end{split}
\end{align}
and
\begin{align}
\begin{split}
16 G_4 (r) = &-36 \log (2) - 3 \pi^2 + 54 (r - \log (2))^2  + 18 \left[2 \ \mathrm{tanh}^2 (r) - 1 \right] r^2 \\ &+3 \ \mathrm{tanh}(r) \Big\{ \pi^2 + 36 (1 - \log (2)) r + 24 r^2 + 12 \mathrm{Li}_2 \left( \mathrm{-e}^{-2r} \right) \Big\} \\ &+72 \Big\{ \mathrm{Li}_2 \left( 1 - \mathrm{cosh}(r) \right) - \mathrm{Li}_2 \left( - \mathrm{cosh}(r) \right) \Big\} \\ &- 36 \log \left( \mathrm{cosh}(r) \right) \Bigg\{ 3 - 2 \log (2) + 3 r + 2 r \ \mathrm{tanh} (r) \\ &\qquad\qquad\qquad\qquad + \log \Big[ \mathrm{sech} (r) \left[ 1 + \mathrm{cosh} (r) \right]^2 \left[ 1 - \mathrm{tanh} (r) \right] \Big] \Bigg\}, \\
16 F_4 (r) = &- 36 \log (2) + 54 (r - \log (2))^2 + 36 r \ \mathrm{tanh} (r) \\ &+ 36 \Big[ \mathrm{Li}_2 \left( 1 - \mathrm{cosh}(r) \right) - \mathrm{Li}_2 \left( - \mathrm{cosh}(r) \right) \Big] \\ &-18 \log \left( \mathrm{cosh} (r) \right) \Bigg\{ 2 (1 - \log (2) + r) + \log \Big[ \mathrm{sech} (r) \left[ 1 + \mathrm{cosh} (r) \right]^2 \Big] \Bigg\}, \\
16 H_4(r) = &72 - 6 \pi^2 + 36 r^2 - 
36 \log \left( \mathrm{cosh} (r) \right) \log \Big[\left[ 1 + \mathrm{cosh} (r) \right] \left[ 1 + \mathrm{coth} (r) \right]^2 \Big] \\ &-\frac{6}{\mathrm{sinh}(2r)} \Big\{\pi^2 + 24 (1 - \log (2)) r + 12 \left[ 1 + \mathrm{tanh}(r) \right] r^2 + 12 \mathrm{Li}_2 \left( - \mathrm{e}^{-2r} \right) \Big\} \\ &+18 \Bigg\{ \mathrm{Li}_2 \left( \mathrm{e}^{-4r} \right) - 4 \mathrm{Li}_2 \left( \mathrm{e}^{-2r} \right)+ 4 \mathrm{Li}_2 \left( \mathrm{e}^{-r} \mathrm{cosh} (r) \right) \\ &\qquad\qquad + 2 \mathrm{Li}_2 \left( 1-\mathrm{cosh} (r) \right) - 2 \mathrm{Li}_2 \left( - \mathrm{cosh}(r) \right) \Bigg\}.
\end{split}
\end{align}

\begin{figure}[H]
\includegraphics[width=9cm]{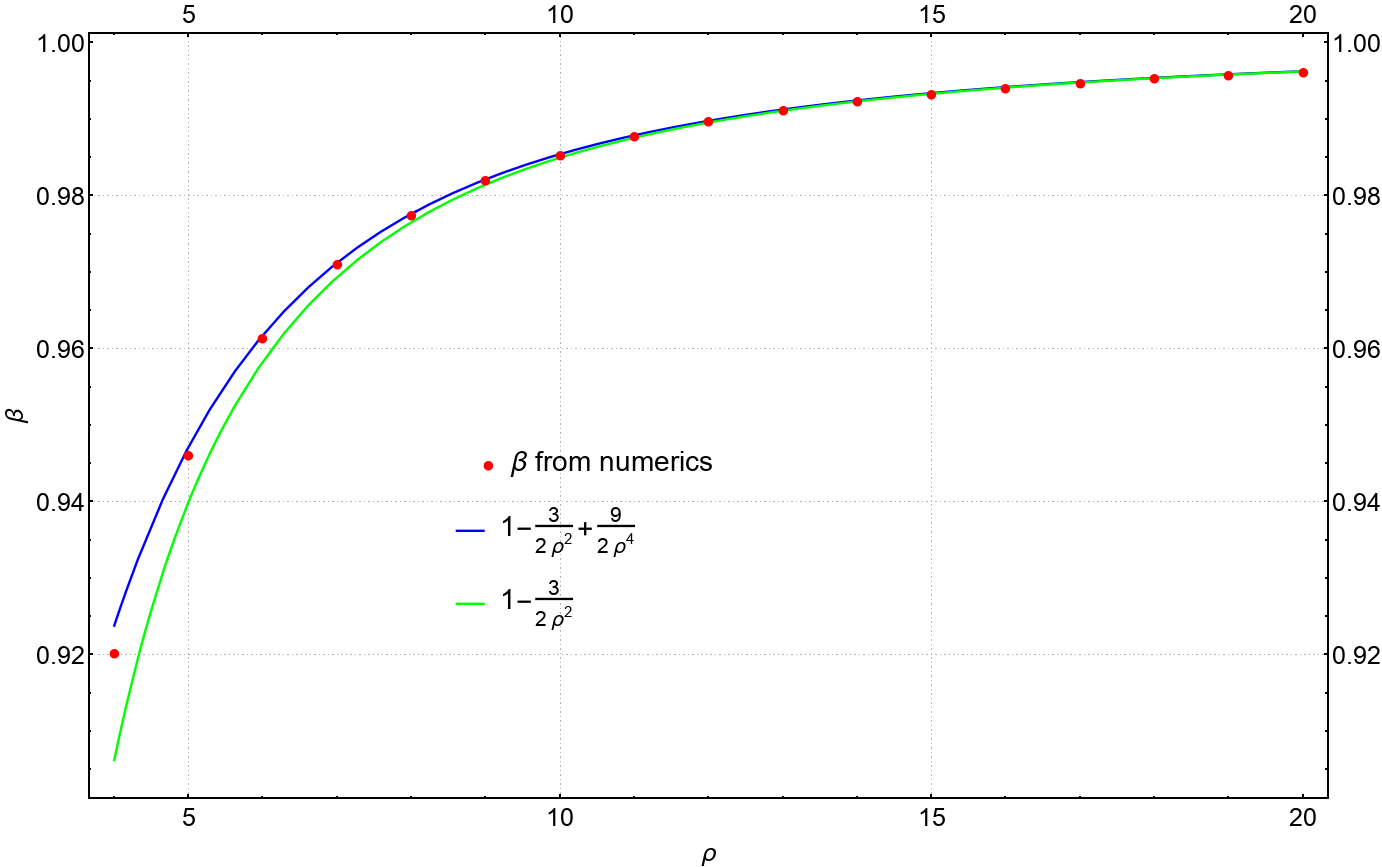} \centering \caption{The circumference $\beta (\rho)$ for large $\rho$, in units of $\beta_H$.} \label{fig:Betarhocomparison}
\end{figure}

All the functions $G_i(r)$, $F_i(r)$, and $H_i(r)$ are everywhere bounded, so the expansion in $1/\rho$ makes sense. The leading behavior of these functions when $r \rightarrow \infty$ is the following:
\begin{align}
\begin{split}
G_2  \approx - \frac{3}{2} \log (2), \quad F_2 &\approx \frac{3}{2} \mathrm{e}^{- 2 r}, \quad H_2 \approx - \frac{3}{2}, \\
G_4 \approx \frac{9}{8} \log (2) \left( 4 - \log (2) \right), \quad F_4&\approx - \frac{9}{4} \left( 2 \log (2) - 1\right) \mathrm{e}^{-2 r}, \quad H_4 \approx \frac{9}{2}.
\end{split}
\end{align}
We can check the asymptotic values of $H_2$ and $H_4$ by comparing them with the numerical data for the asymptotic radius of the $S^1$ as a function of $\rho$, see Figure
\ref{fig:Betarhocomparison}.

Our solutions make sense for any $f(\rho)$, but in order to compare different solutions we need them to have the same asymptotic behavior. One can check that if we choose the normalization of the string coupling to be
\begin{equation}
f(\rho) = g_0 \mathrm{e}^{-\rho^2/3} \left\{ 1 + \frac{\pi^2/8}{\rho^2} - \frac{\frac{9}{4} \log (2) \left( 2 - \log (2) \right) - \frac{\pi^4}{128} + \frac{A_4}{3}}{\rho^4}+ \mathcal{O}\left(\frac{1}{\rho^6}\right)\right\}
\end{equation}
with constant $g_0$, then we have
\begin{equation}
g(\zeta_0 (r)) \approx g_0 \zeta_0 (r) \mathrm{e}^{ - \zeta_0 (r)^2/3}, \label{dilnorm}
\end{equation}
which exactly coincides with (\refeq{gzetasingular}). This will allow us to compare the black hole solution to the solution describing a thermal gas in the LST on $S^4$ (which is given by \eqref{exacts1} with a constant radius of the $S^1$).

Next we study the thermal properties of our solution. The temperature $T$ is determined from the requirement of correct periodicity of the compact direction in the string frame, and is given by
\begin{equation}
T = \left[2 \pi \sqrt{N} \left( 1 - \frac{3/2}{\rho^2} + \frac{9/2}{\rho^4} + \mathcal{O} \left(\frac{1}{\rho^6}\right)\right)\right]^{-1} = \frac{1}{\beta_H} \left[ 1 + \frac{3/2}{\rho^2} - \frac{9/4}{\rho^4} +\mathcal{O}\left(\frac{1}{\rho^6}\right)\right].
\end{equation}

The area of the black hole horizon in string units is
\begin{equation}
\mathrm{A} = N^{7/2} V_{S^3} V_{S^4} \zeta_0(0)^4 V_{\mathbb{R}} = N^{7/2} V_{S^3} V_{S^4} V_{\mathbb{R}} \rho^4,
\end{equation}
where $V_{S^3}$ and $V_{S^4}$ are the volumes of the unit 3- and 4-spheres respectively, and $V_{\mathbb{R}}$ is the volume of the non-compact direction (which contains a factor of $\sqrt{N}$). The entropy is of the form
\begin{equation} \label{Entropy}
S = \frac{\mathrm{A}}{4 G_N}=\frac{4 \pi \mathrm{A}}{(2 \pi)^7 g(0)^2} = \chi \rho^2 \left\{ 1 - \frac{\pi^2/4}{\rho^2} + \frac{\frac{9}{2} \log (2) \left( 2 - \log (2) \right) + \frac{\pi^4}{32} + \frac{2 A_4}{3}}{\rho^4} +\mathcal{O}\left(\frac{1}{\rho^6}\right) \right\} \mathrm{e}^{2 \rho^2/3},
\end{equation}
where we introduced the notation $\chi \equiv \frac{4 \pi N^{7/2} V_{S^3} V_{S^4} V_{\mathbb{R}}}{(2 \pi)^7 g_0^2}$.

In the canonical ensemble
\begin{equation}
\beta_H \mathrm{d} E = \frac{T}{T_H} \mathrm{d}S = \left[ 1 + \frac{3/2}{\rho^2} - \frac{9/4}{\rho^4} +\mathcal{O}\left(\frac{1}{\rho^6}\right)\right] \mathrm{d} S.
\end{equation}
Integrating this relation, we get
\begin{equation} \label{Energy}
\beta_H E = \chi \rho^2 \left\{ 1 - \frac{\frac{\pi^2}{4}-\frac{3}{2}}{\rho^2} + \frac{\frac{9}{2} \log (2) \left( 2 - \log (2) \right) - \frac{\pi^2}{32} (12 - \pi^2) + \frac{2 A_4}{3}}{\rho^4} + \mathcal{O} \left(\frac{1}{\rho^6}\right)\right\} \mathrm{e}^{2 \rho^2/3},
\end{equation}
and the free energy is
\begin{equation}
F = E - T S = \frac{9}{4} \frac{\chi}{\beta_H} \mathrm{e}^{2 \rho^2/3} \left[ \frac{1}{\rho^2}+ \mathcal{O} \left(\frac{1}{\rho^4}\right) \right]. \label{FEhagedorn}
\end{equation}
We know that in the canonical ensemble the following relation must hold:
\begin{equation}
S = \beta^2 \frac{\partial F}{\partial \beta} \simeq \frac{2}{3} \beta_H \rho^4 \frac{\partial F}{\partial \rho^2},
\end{equation}
and indeed we can see that it is satisfied at leading order in the $1/\rho^2$ expansion. To go beyond the leading order, we must compute the entropy and energy to orders higher than $1/\rho^4$.

We can also find $S(E)$. For convenience we introduce the entropy and energy densities $\sigma = S/ V_{\mathbb{R}}$ and $\epsilon = E/ V_{\mathbb{R}}$, and obtain (in our $\alpha'=1$ units)
\begin{equation} \label{Equationofstate}
\beta_H \epsilon = \sigma \left[ 1 + \frac{1}{\log (\sigma)} + \frac{\log \left( \log (\sigma) \right)}{\log^2 (\sigma)} + o \left( \frac{\log \left( \log (\sigma) \right)}{\log^2 (\sigma)} \right) \right],
\end{equation}
or
\begin{equation} \label{Equationofstate2}
\sigma = \beta_H \epsilon \left[ 1 - \frac{1}{\log \left( \beta_H \epsilon \right)} - \frac{\log \left( \log \left( \beta_H \epsilon \right) \right)}{\log^2 \left( \beta_H \epsilon \right)} + o \left( \frac{\log \left( \log \left( \beta_H \epsilon \right) \right)}{\log^2 \left( \beta_H \epsilon \right)} \right) \right].
\end{equation}
We see that the leading correction doesn't depend on any parameter of our geometry (neither $N$ nor $g_0$). This thermodynamical behavior is very similar to the one found in \cite{BuchelInstability}.
As expected from the form of the classical solutions, the corrections to the Hagedorn behavior are much larger at high energies than in the flat space case \eqref{hagcorr}.

\begin{figure}[H]
\includegraphics[width=9cm]{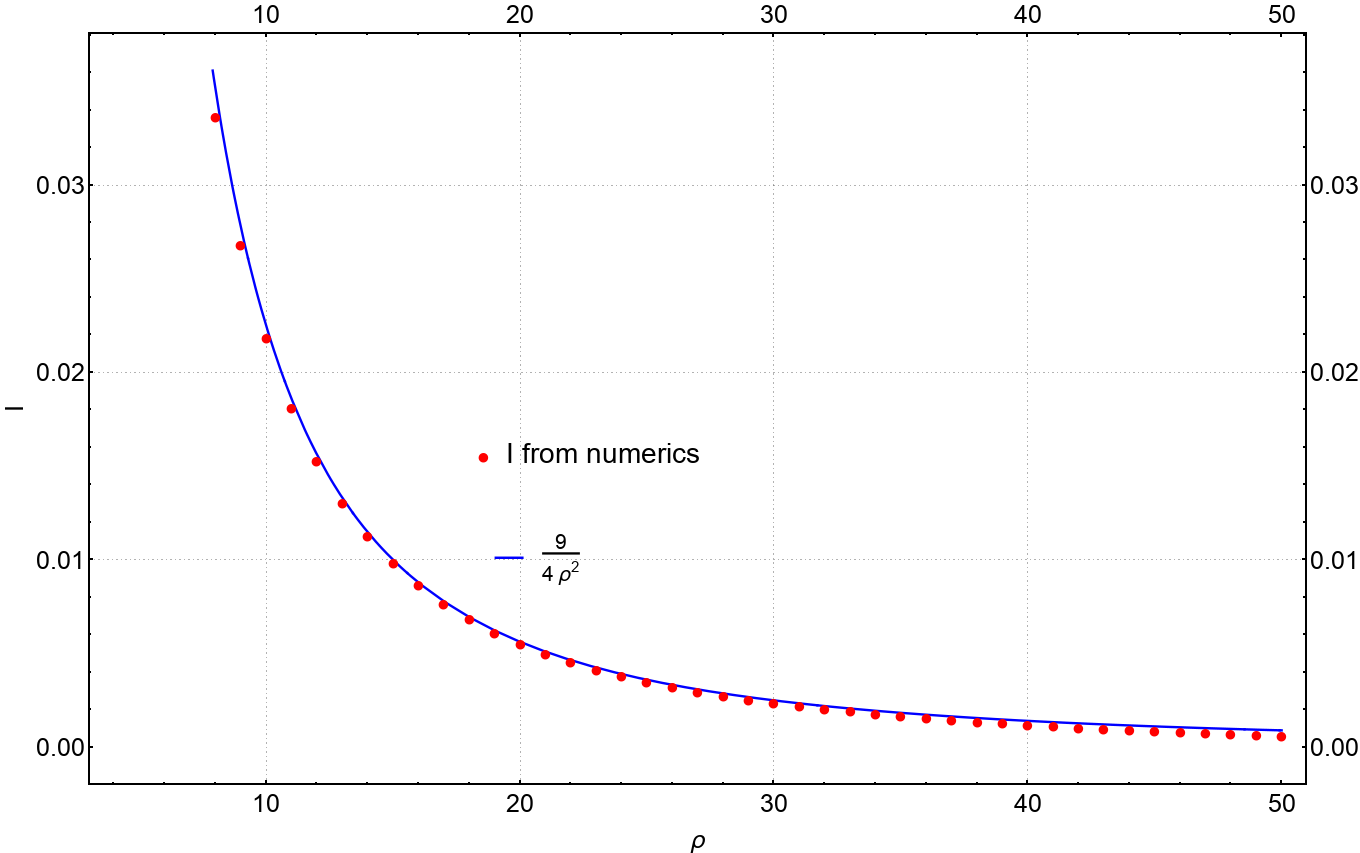} \centering \caption{The dependence of $I (\rho)\mathrm{e}^{-2\rho^2/3} / \chi$ on $\rho$ for large $\rho$, compared to the expectation \eqref{FEhagedorn}.}
\label{fig:Actionrho}
\end{figure}

The behavior \eqref{FEhagedorn} can be seen also from the action $I = \beta F$, which we can compute for our numerical solutions using the procedure outlined in the beginning of the section.
It is easy to check that all terms in the action are proportional to $\chi$. In Figure \ref{fig:Actionrho} we plot the action evaluated in this way, divided by $\chi$ and multiplied by $\mathrm{e}^{-2\rho^2/3}$, for various values of $\rho$, and we see that it agrees well with the leading order term in \eqref{FEhagedorn} (note that at leading order in $1/\rho$, $\beta = \beta_H$).

\begin{figure}[H]
\includegraphics[width=9cm]{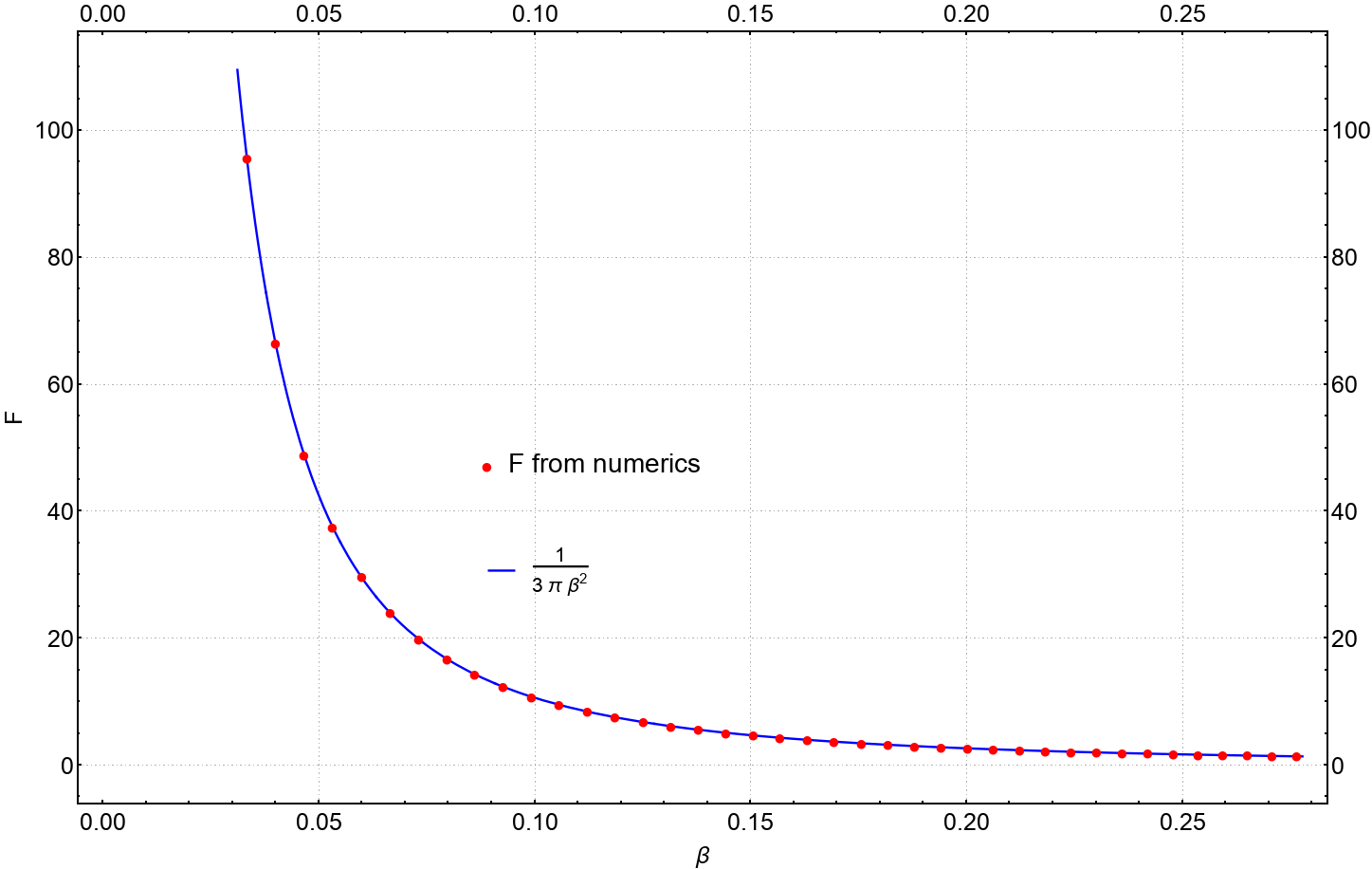} \centering \caption{$F(\beta)$ for the black hole solutions in the small-$\beta$ region, in units of $\chi$, compared to $\frac{\beta_H^2}{3\pi \beta^2}$. Here $\beta$ is measured in units of $\beta_H$.} \label{fig:fesmallbeta}
\end{figure}

We can also consider small values of $\rho$. For these values of $\rho$ we cannot use the large-$\rho$ expansion, but we are able to compute the action (and the free energy) numerically. We find numerically that in this regime $\beta / \beta_H = 2 \rho / 3 + \mathcal{O} (\rho^2)$,  and that the free energy approximately behaves as $\frac{3 \chi}{4\pi\rho^2}$ (or as $\frac{\chi \beta_H^2}{3\pi \beta^2}$), see Figures \ref{fig:fesmallrho} and \ref{fig:fesmallbeta}:

\begin{figure}[H]
\includegraphics[width=9cm]{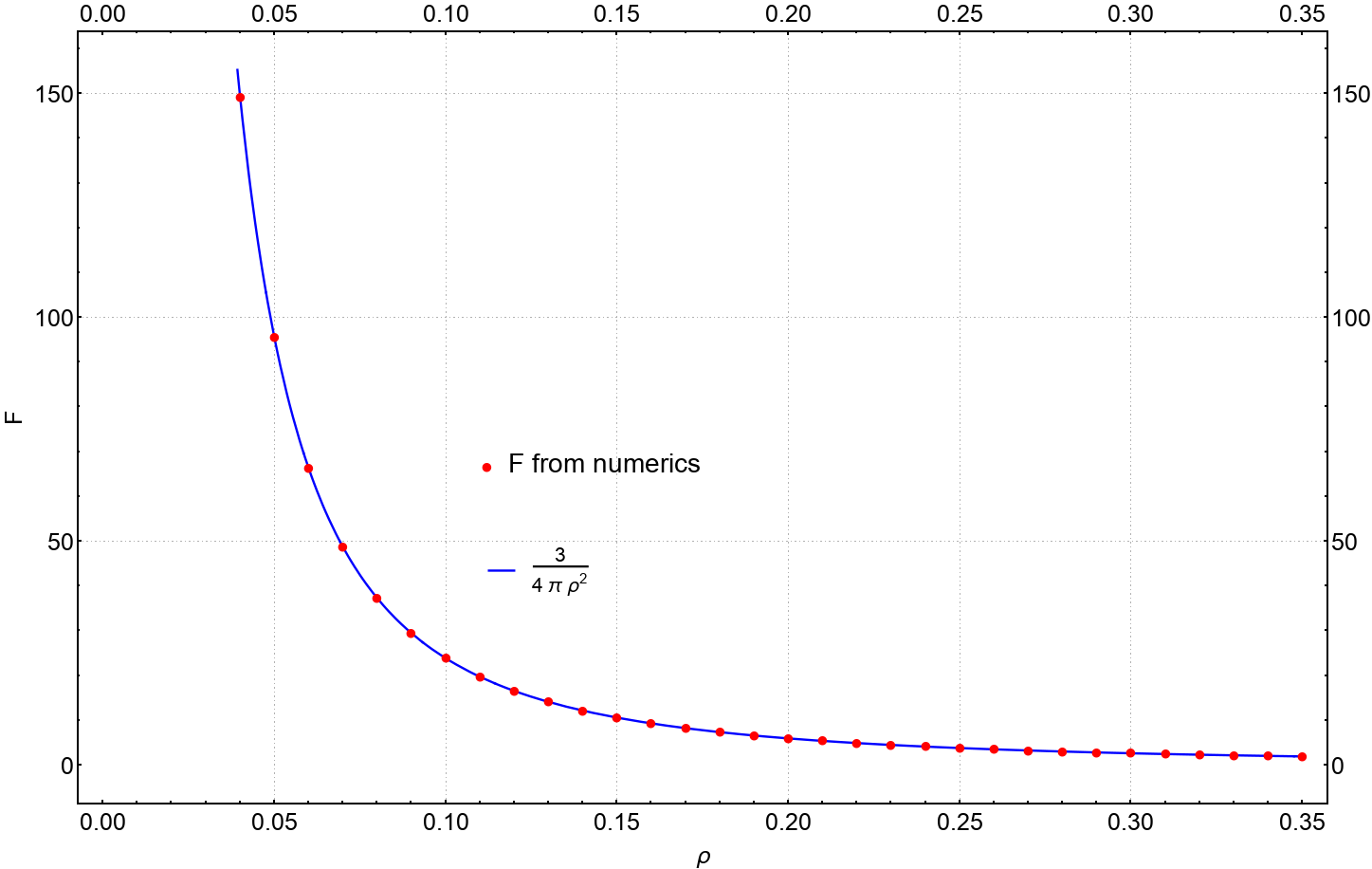} \centering \caption{$F(\rho)$ for the black hole solutions in the small-$\rho$ region, in units of $\chi$, compared to $\frac{3}{4\pi\rho^2}$.} \label{fig:fesmallrho}
\end{figure}

In addition to the black hole solutions that we discussed up to now, we also have, for any temperature $\beta$, another solution with the same asymptotics, in which the $S^4$ vanishes at $r=0$ and not the $S^1$. This solution is simply \eqref{exacts1}, with a constant radius for the $S^1$ as a function of the radial direction (in the string frame). For this solution we can analytically  compute the action and the free energy, and we  find that for any $\beta$
\begin{equation}
I = \frac{3}{\pi\mathrm{e}^2 \sqrt{N}} \chi\beta, \qquad F = \frac{3}{\pi\mathrm{e}^2 \sqrt{N}} \chi.
\end{equation}

\begin{figure}[H]
\includegraphics[width=9cm]{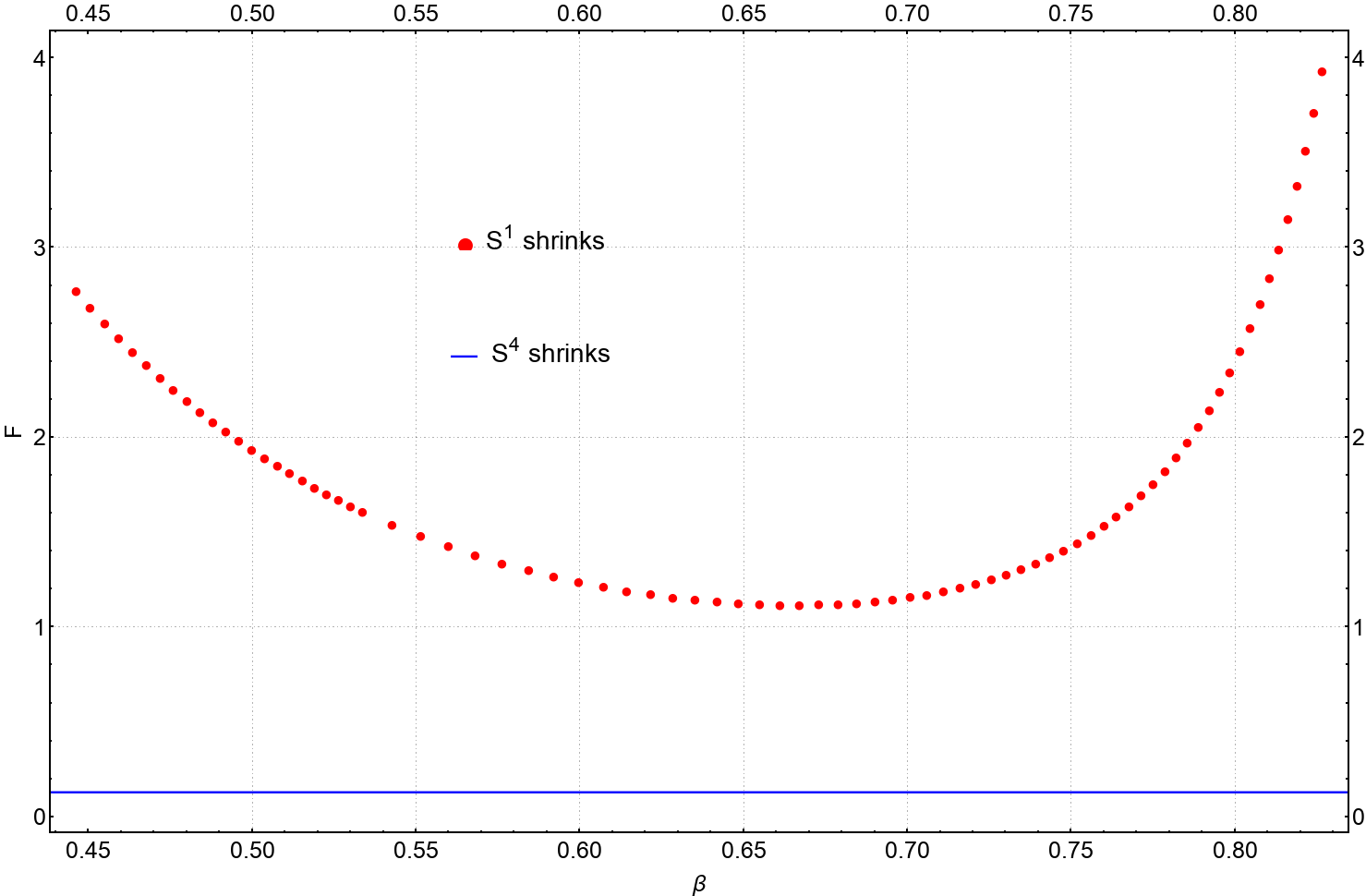} \centering \caption{$\sqrt{N} F(\beta)/\chi$ for the solutions with shrinking $S^1$, $S^4$ warp factors. $\beta$ is measured in units of $\beta_H$.} \label{fig:fes1s4b}
\end{figure}

In the range of temperatures $T > T_H$ ($\beta < \beta_H$) where both solutions exist, we can compare their free energies. Note that we defined $g_0$ in the solutions \eqref{exacts1} so that it will agree with the asymptotics \eqref{dilnorm}, enabling a direct comparison of the two solutions. We find that the solution with the $S^4$ shrinking  has a smaller free energy for all values of $\beta$, and is therefore preferred (see Figure \ref{fig:fes1s4b}). However, given the Hagedorn behavior, the thermal ensemble in any case does not exist for $T > T_H$, so the meaning of this observation is not clear.

\subsection{Instability}

The temperatures of all the black hole solutions we found are above the Hagedorn temperature, so we expect the canonical ensemble to be ill-defined, and the corresponding solutions to be unstable. Indeed, it is easy to compute the specific heat at large $\rho$ from (\refeq{Energy}), and to find that it is negative (at leading order in $1/\rho$):
\begin{equation}
c_V = \frac{\mathrm{d} E}{\mathrm{d} T} \approx - \frac{2}{3} \beta_H \rho^4 \frac{\mathrm{d} E}{\mathrm{d} \rho^2} \approx - \frac{4}{9} \chi \rho^6 \mathrm{e}^{2 \rho^2/3}.
\end{equation}

According to the Gubser-Mitra conjecture \cite{GubserMitra}, gravitational backgrounds with a translationally invariant horizon, corresponding to a black brane geometry, develop an instability (a tachyonic mode) precisely when the specific heat of the black brane becomes negative. Our solution is exactly of this type with the topology of the horizon being $\mathbb{R} \times S^3 \times S^4$. An instability of the gravitational solution gets mapped to a field theoretic instability \cite{BuchelInstability} under the gauge/gravity duality \cite{MaldacenaHol, GKPHol, WittenHol}, manifesting itself as an imaginary velocity of sound. One can argue this as follows. The sound velocity in a medium can be computed as
\begin{equation}
v_s^2 = \frac{\partial \mathcal{P}}{\partial \mathcal{E}},
\end{equation}
where $\mathcal{P}$ and $\mathcal{E}$ are the pressure and the energy density, respectively. At zero chemical potential and fixed volume $V$, this becomes
\begin{equation}
v_s^2 = \frac{\left( \partial \mathcal{P} / \partial T \right)_V}{\left( \partial \mathcal{E} / \partial T \right)_V} = \frac{S}{c_V},
\end{equation}
which is negative for negative $c_V$, because the entropy (corresponding to the horizon area on the gravity side of the duality) is positive.
In particular, 
we can compute the velocity of sound for our solution, and get
\begin{equation}
v_s^2 = \frac{S}{c_V} = - \frac{9}{4\rho^4} + \mathcal{O} \left( \frac{1}{\rho^6} \right).
\end{equation}

\section{Oscillations near singular solutions} \label{oscs}

Whenever we have two curved compact factors, and one of them shrinks to zero, the other one has some fixed size $\rho$ at the origin; we can take the limit $\rho \to 0$ to obtain a solution where both factors shrink at the same time. Such a solution is always singular at $r=0$, and locally near $r=0$ the metric is proportional to 
\begin{equation} \label{con_sing}
\mathrm{d} r^2 + \frac{n_1-1}{n_1+n_2-1}r^2 \mathrm{d} \Omega_{n_1}^2 + \frac{n_2-1}{n_1+n_2-1} r^2 \mathrm{d} \Omega_{n_2}^2,
\end{equation}
where the compact factors are $S^{n_1}$ and $S^{n_2}$ ($n_1,n_2 \geq 2$). Singular solutions of this type arose also in other contexts, such as the black hole/black string phase transition  \cite{Kol, CriticalBehavior, CardonaFigueras} and the holographic duals of CFTs on products of spheres \cite{Erez}. The expansion in small fluctuations around \eqref{con_sing} in flat space may be done analytically, and for $n_1+n_2 < 9$ the leading fluctuations have oscillatory behavior as a function of $r$. For $r\ll 1$ our solutions look approximately like they are in flat space, which means that if we turn on some small size $\rho \ll 1$ for one of the compact spaces, then in the region $\rho \ll r \ll 1$ this will turn on a fluctuation of the singular solution \eqref{con_sing} that looks just like a small fluctuation around the flat space solution, and will thus oscillate as a function of $r$. If we consider the solution at a fixed value of $r$, then the values of the various metric components will then exhibit oscillations as a function of $\rho$ \cite{Kol, CriticalBehavior, CardonaFigueras, Erez}.

The full solutions for all $r$ will be very different than the flat space ones, but for $\rho \ll 1$ the full solutions will exhibit similar oscillations as a function of $\rho$, since we have the oscillations in the small $r$ region, and the full solutions can be found by starting from the oscillating region and integrating from there to larger values of $r$. In particular, we expect that also the asymptotic parameters that we find will exhibit oscillations as a function of $\rho$, around the values of these parameters that arise in the singular solution with $\rho=0$. In this section we exhibit these oscillations for several examples. The oscillations imply that for values of the asymptotic parameters close to the critical ones corresponding to $\rho=0$, there are several different solutions with the same topology for each value of the asymptotic parameters.


\subsection{$\mathcal{M} = S^3 \times S^3$} \label{s3s3osc}

We start with a singular solution with both spheres shrinking to zero, for which the corresponding warp factors $\zeta_0$ and $\zeta_1$, and the dilaton $g$, have the form
\begin{align}
\begin{split} \label{sing_sol}
\zeta_0 \left( r \right) &= \zeta_1 \left( r \right) = \sqrt{\frac{2}{5}} r + \mathcal{O} \left( r^3 \right), \\ g \left( r \right) &= 1 + \mathcal{O} \left( r^2 \right).
\end{split}
\end{align}
Clearly the solution for this case will have $\zeta_0=\zeta_1$ for all $r$, and also the asymptotic parameters will obey $\alpha_{1,0}^0=\alpha_{1,0}^1$.

We can now deform this continuously to a solution with non-zero $\zeta_0(0) = \rho \ll 1$. We expect that for small $\rho$ this will significantly change the solution around $r \sim \rho$, but that the solution for $\rho \ll r \ll 1$ will just be a small correction to \eqref{sing_sol}. We can find the most general form of the corrections in this region by considering linearized fluctuations around \eqref{sing_sol}; in the region $\rho \ll r \ll 1$ we can just keep the leading term in \eqref{sing_sol}, such that we are simply expanding around a conical singularity in flat space.
The equations are consistent with a linearized fluctuation obeying $\delta g = 0$ and $\delta \zeta_0 = - \delta \zeta_1 = \mathfrak{f}$, and they reduce to a single equation for $\mathfrak{f}$:
\begin{equation}
\frac{\mathrm{d}^2 \mathfrak{f}}{\mathrm{d} r^2} + \frac{4}{r} \frac{\mathrm{d} \mathfrak{f}}{\mathrm{d} r} + \frac{6}{r^2} \mathfrak{f} = 0.
\end{equation}
This is an Euler equation whose general solution is
\begin{equation} \label{lin_sol}
\mathfrak{f} \left( r \right) = \frac{Z}{r^{3/2}} \sin \left( \frac{\sqrt{15}}{2} \log (r) + \phi_0 \right).
\end{equation}

Numerically computing $\zeta_0(r)-\zeta_1(r)$ for small $\rho$ indeed gives a solution in the range $\rho \ll r \ll 1$ that takes this form,
see Figure \ref{fig:S3S3solution}):
\begin{figure}[H]
\includegraphics[width=11cm]{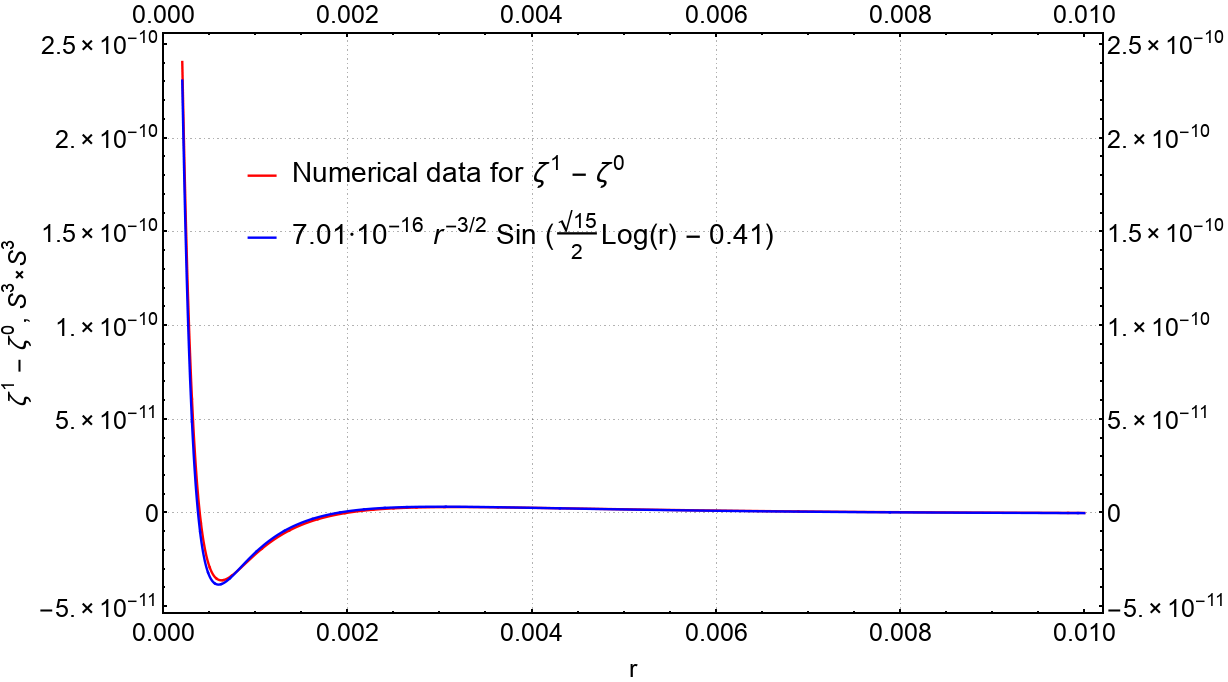} \centering \caption{The warp factor difference fit $\zeta_1 - \zeta_0$ for $\mathcal{M} = S^3 \times S^3$ with $\rho = 10^{-6}$.} \label{fig:S3S3solution}
\end{figure}
 
In general the parameters $Z$ and $\phi_0$ in \eqref{lin_sol} depend on $\rho$. We can fix the dependence of $Z$ on $\rho$ at small $\rho$ by dimensional analysis, since the relative change in the solutions in the range of interest (namely, $\delta \zeta_0 / \zeta_0$ and $\delta \zeta_1 / \zeta_1$) is invariant under rescaling $r$ and $\rho$ together. As in \cite{Erez}, this implies that at leading order in $\rho$
\begin{equation}
Z(\rho) \propto \rho^{5/2} \sin \left( \frac{\sqrt{15}}{2} \log (\rho) + \varphi_0 \right),
\end{equation}
for some constant $\varphi_0$.

\begin{figure}[H]
\includegraphics[width=11cm]{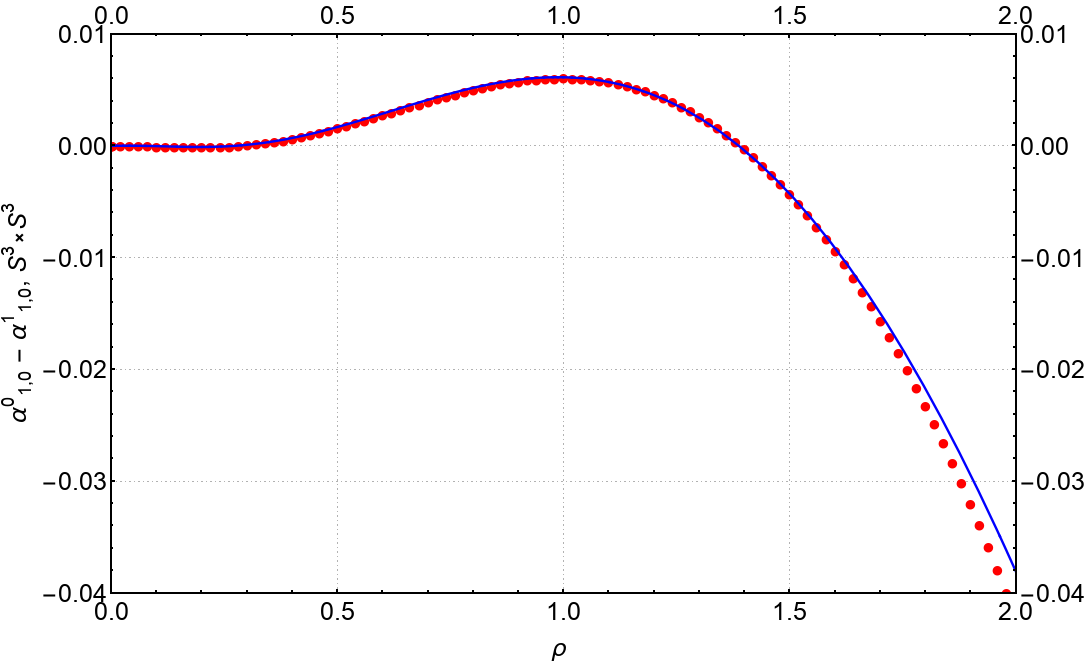} \centering \caption{A fit of $\alpha_{1,0}^0 - \alpha_{1,0}^1$, as a function of $\rho$, to \eqref{expdiff} for $\mathcal{M} = S^3 \times S^3$.} \label{fig:S3S3coefficients}
\end{figure}

If we now consider the solution for the full range of $r$, obtained by integrating the solution from the region where \eqref{lin_sol} is valid towards larger values of $r$, then its leading deviation from the $\rho=0$ solution will be proportional to $Z(\rho)$, and thus we expect also the asymptotic parameters of our solutions to change in a way that is proportional to this. As discussed in section \ref{sec:asymp} the natural asymptotic parameter in this case is $\alpha_{1,0}^0-\alpha_{1,0}^1$, and we indeed find numerically that for small $\rho$
(see  Figure \ref{fig:S3S3coefficients})
\begin{equation} \label{expdiff}
\alpha^0_{1,0}(\rho) - \alpha^1_{1,0}(\rho) \approx 0.0103 \rho^{5/2} \sin \left( {\frac{\sqrt{15}}{2} \log(\rho)} + 5.650 \right).
\end{equation}

We see from the figure that for some values of the asymptotic parameter $\alpha_{1,0}^0-\alpha_{1,0}^1$ there is more than one solution, with different values of $\rho$. The oscillations become faster and faster for small $\rho$, such that as $|\alpha_{1,0}^0-\alpha_{1,0}^1|$ becomes smaller and smaller the number of solutions grows, and for $\alpha_{1,0}^0-\alpha_{1,0}^1=0$ there is actually an infinite number of different solutions (converging to the singular solution with $\rho=0$).\footnote{This is true in the supergravity approximation. For small enough $\rho$ (of order $1/\sqrt{N}$) stringy corrections start becoming important, and they may modify this behavior.} A similar behavior appears already for conformal field theories on products of spheres \cite{Erez}, so that it is not specific to LSTs. When we have several solutions, we can have phase transitions between them as we change the asymptotic parameter $(\alpha_{1,0}^0-\alpha_{1,0}^1)$ (as analyzed for CFTs on products of spheres in \cite{Erez}); to analyze this we would need to compute the free energies of the different solutions, but as explained in section \ref{renorm}, we are not yet able to do this.

\subsection{$\mathcal{M} = \mathbb{R}^2 \times S^2 \times S^2$}

The situation here is very similar to what we discussed in the previous subsection. The singular solution around $r=0$ with two shrinking $S^2$'s has the warp factors and the dilaton
\begin{align}
\begin{split}
\tilde{\zeta}_0 \left( r \right) &= \tilde{\zeta}_1 \left( r \right) = \frac{1}{\sqrt{3}} r + \mathcal{O} \left( r^3 \right), \\ \tilde{\zeta}_3 \left( r \right) &= 1 + \mathcal{O} \left( r^2 \right), \\ \tilde{g} \left( r \right) &= 1 + \mathcal{O} \left( r^2 \right).
\end{split}
\end{align}
Here $\tilde{\zeta}_0$ and $\tilde{\zeta}_1$ correspond to the two $S^2$'s, and $\tilde{\zeta}_3$ is the $\mathbb{R}^2$ warp factor; again symmetry implies that the full singular solution has $\tilde{\zeta}_0=\tilde{\zeta}_1$, and $\tilde{\alpha}_{1,0}^0 = \tilde{\alpha}_{1,0}^1$.

\begin{figure}[H]
\includegraphics[width=11cm]{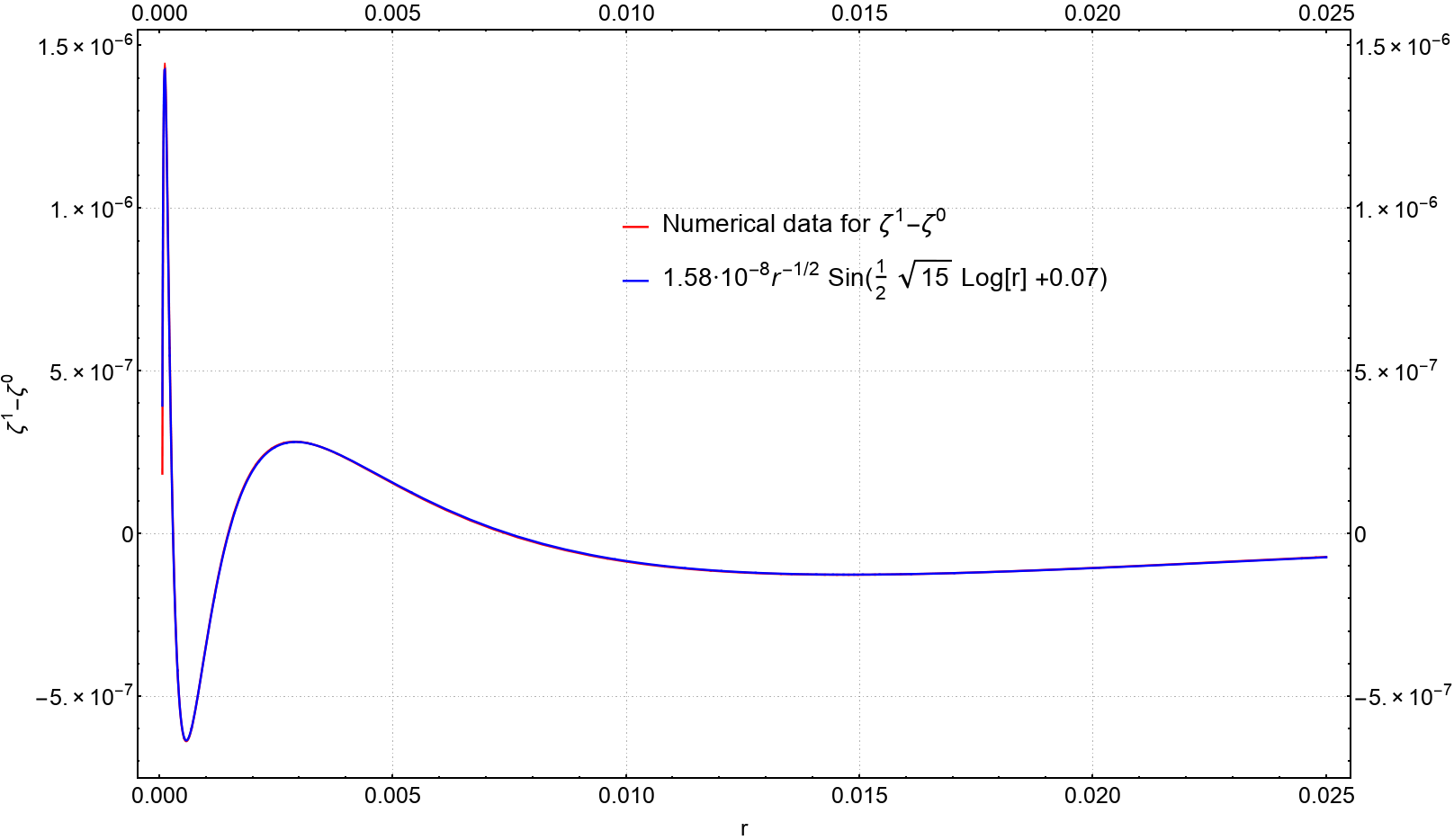} \centering \caption{The warp factor difference fit for $\tilde{\zeta}_1 - \tilde{\zeta}_0$ for $\mathcal{M} = \mathbb{R}^2 \times S^2 \times S^2$, $\tilde{\rho} = 10^{-5}$.} \label{fig:R2S2S2solution}
\end{figure}

A consistent ansatz for small perturbations of this solution is $\delta \tilde{g} = \delta \tilde{\zeta}_3 = 0$, and $\delta \tilde{\zeta}_0 = - \delta \tilde{\zeta}_1 = \tilde{\mathfrak{f}}$. The equations of motion again reduce to one equation on $\tilde{\mathfrak{f}}$ of the form
\begin{equation}
\frac{\mathrm{d}^2 \tilde{\mathfrak{f}}}{\mathrm{d} r^2} + \frac{2}{r} \frac{\mathrm{d} \tilde{\mathfrak{f}}}{\mathrm{d} r} + \frac{4}{r^2} \tilde{\mathfrak{f}} = 0.
\end{equation} 
The solution is
\begin{equation}
\mathfrak{\tilde{f}} \left( r \right) = \frac{\tilde{Z}}{\sqrt{r}} \sin \left( \frac{\sqrt{15}}{2} \log (r) + \tilde{\phi}_0 \right),
\end{equation}
and it agrees well with our numerical solutions when we turn on a small non-zero radius $\tilde{\zeta}_0(0) = \tilde{\rho}$, see Figure \ref{fig:R2S2S2solution}.

As in the previous subsection, general arguments imply that for small ${\tilde \rho}$ we should have
\begin{equation}
{\tilde Z}(\tilde \rho) \propto \tilde{\rho}^{3/2} \sin \left( {\frac{\sqrt{15}}{2} \log({\tilde{\rho}})} + \tilde{\varphi}_0 \right).
\end{equation}
We can indeed fit the numerical solutions for $(\tilde{\alpha}_{1,0}^0 - \tilde{\alpha}_{1,0}^1)$ at various small values of ${\tilde \rho}$ with this form:
\begin{equation} \label{forfig16}
	\tilde{\alpha}^0_{1,0}(\tilde{\rho}) - \tilde{\alpha}^1_{1,0}(\tilde{\rho}) \approx -0.0802 \tilde{\rho}^{3/2} \sin \left( {\frac{\sqrt{15}}{2} \log({\tilde{\rho}})} + 1.779 \right),
\end{equation}
see Figure
%
\ref{fig:R2S2S2coefficients}:

\begin{figure}[H]
\includegraphics[width=11cm]{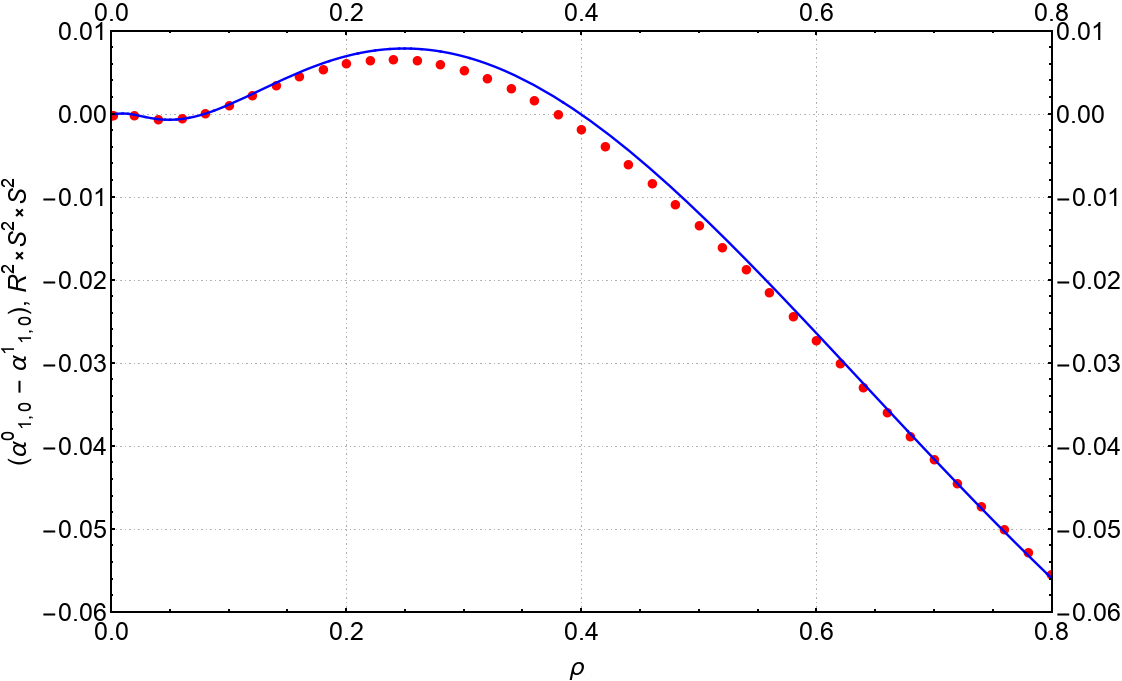} \centering \caption{A fit of $\tilde{\alpha}_{1,0}^0 - \tilde{\alpha}_{1,0}^1$ to \eqref{forfig16} for $\mathcal{M} = \mathbb{R}^2 \times S^2 \times S^2$.} \label{fig:R2S2S2coefficients}
\end{figure}

\subsection{$\mathcal{M} = S^2 \times S^4$}

This case is again similar, except that there is no longer a symmetry between the two spheres, so now we can turn on a small radius for one sphere or the other, and the behavior will be different in the two cases. Both will exhibit oscillations, but the coefficients can be different.

The singular solution with both spheres shrinking at the origin is now
\begin{align}
\begin{split}
\hat{\zeta}_4 \left( r \right) &= \sqrt{\frac{3}{5}} r + \mathcal{O} \left( r^3 \right), \\ \hat{\zeta}_2 \left( r \right) &= \frac{1}{\sqrt{5}} r + \mathcal{O} \left( r^3 \right), \\ \hat{g} \left( r \right) &= 1 + \mathcal{O} \left( r^2 \right),
\end{split}
\end{align}
where $\hat{\zeta}_4$ corresponds to $S^4$, and $\hat{\zeta}_2$ corresponds to $S^2$.

The ansatz for small perturbations is now $\delta \hat{g} = 0$, and $\delta \hat{\zeta}_4 = - \frac{\sqrt{3}}{2} \delta \hat{\zeta}_2 = \hat{\mathfrak{f}}$, which gives the equation
\begin{equation}
\frac{\mathrm{d}^2 \hat{\mathfrak{f}}}{\mathrm{d} r^2} + \frac{4}{r} \frac{\mathrm{d} \hat{\mathfrak{f}}}{\mathrm{d} r} + \frac{6}{r^2} \hat{\mathfrak{f}} = 0,
\end{equation}
exactly as we had in the $S^3 \times S^3$ case. The solution is given by
\begin{equation}
\mathfrak{\hat{f}} \left( r \right) = \frac{\hat{Z}}{r^{3/2}} \sin \left( \frac{\sqrt{15}}{2} \log (r) + \hat{\phi}_0 \right),\label{oscsol2}
\end{equation}
and it fits well numerically with the solutions where we turn on small radii, see an example in Figure \ref{fig:S2S4solution}:
\begin{figure}[H]
\includegraphics[width=11cm]{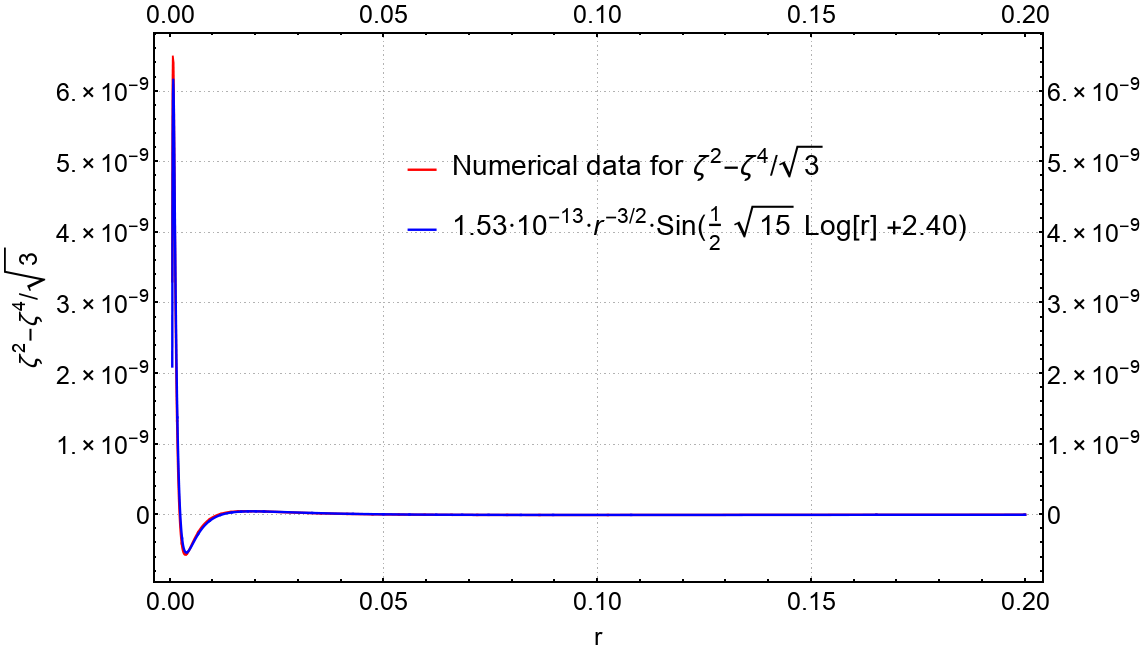} \centering \caption{The warp factor difference $\hat{\zeta}_2 - \hat{\zeta}_4/\sqrt{3}$ for $\mathcal{M} = S^2 \times S^4$, fit to the behavior \eqref{oscsol2}. The value of the non-vanishing radius is $\hat{\zeta}_4(0) = \hat{\rho} = 10^{-5}$.} \label{fig:S2S4solution}
\end{figure}

\begin{figure}[H]
\includegraphics[width=11cm]{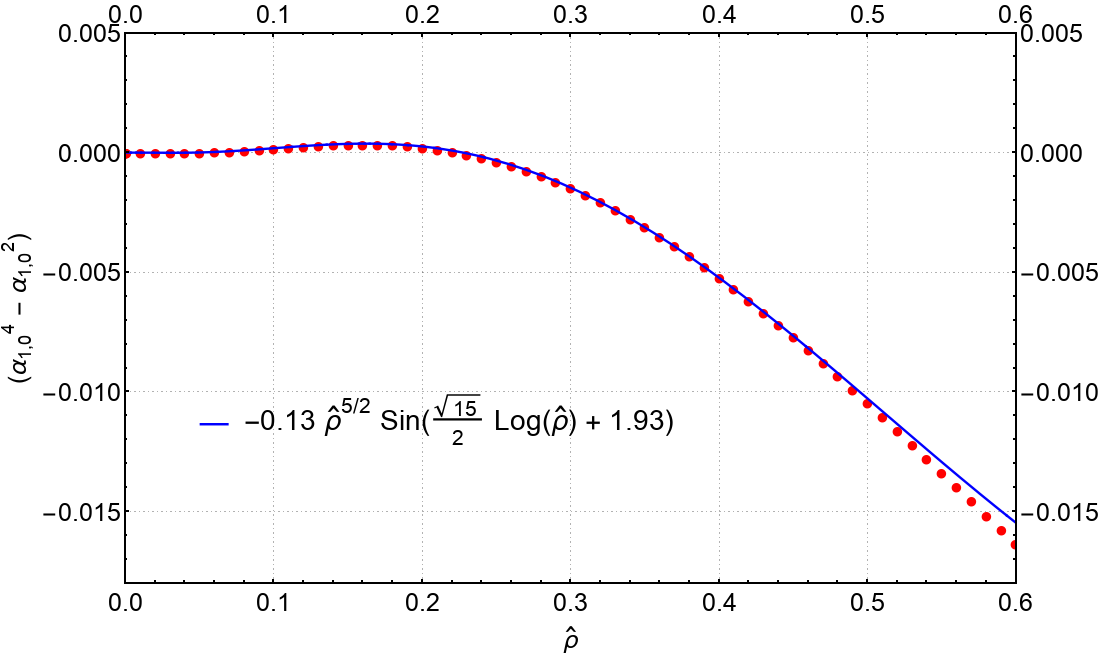} \centering \caption{A fit of $\hat{\alpha}_{1,0}^4 - \hat{\alpha}_{1,0}^2$ for $\mathcal{M} = S^2 \times S^4$ when $S^4$ shrinks.} \label{fig:S2S4coefficientsS4shrinks}
\end{figure}

Once again, when we turn on a small radius $\hat{\rho}$ at the origin either for the $S^2$ or for the $S^4$, we can use scaling to argue that we should have
\begin{equation}
\hat{Z} \propto \hat{\rho}^{5/2} \sin \left( \frac{\sqrt{15}}{2} \log (\hat{\rho}) + \hat{\varphi}_0 \right), \label{coeffosc}
\end{equation}
and we expect to find this behavior also for the asymptotic parameter $(\hat{\alpha}^4_{1,0} - \hat{\alpha}^2_{1,0})$ (here the superscript refers to the dimension of the sphere). And indeed, our numerical solutions for small ${\hat \rho}$ agree with this behavior in both cases, but with different coefficients, see Figure \ref{fig:S2S4coefficientsS2shrinks} for the case where the $S^2$ shrinks and ${\hat \rho}$ is the radius of the $S^4$ at $r=0$, and Figure \ref{fig:S2S4coefficientsS4shrinks} for the case where the $S^4$ shrinks and ${\hat \rho}$ is the radius of the $S^2$ at $r=0$:


\begin{figure}[H]
\includegraphics[width=11cm]{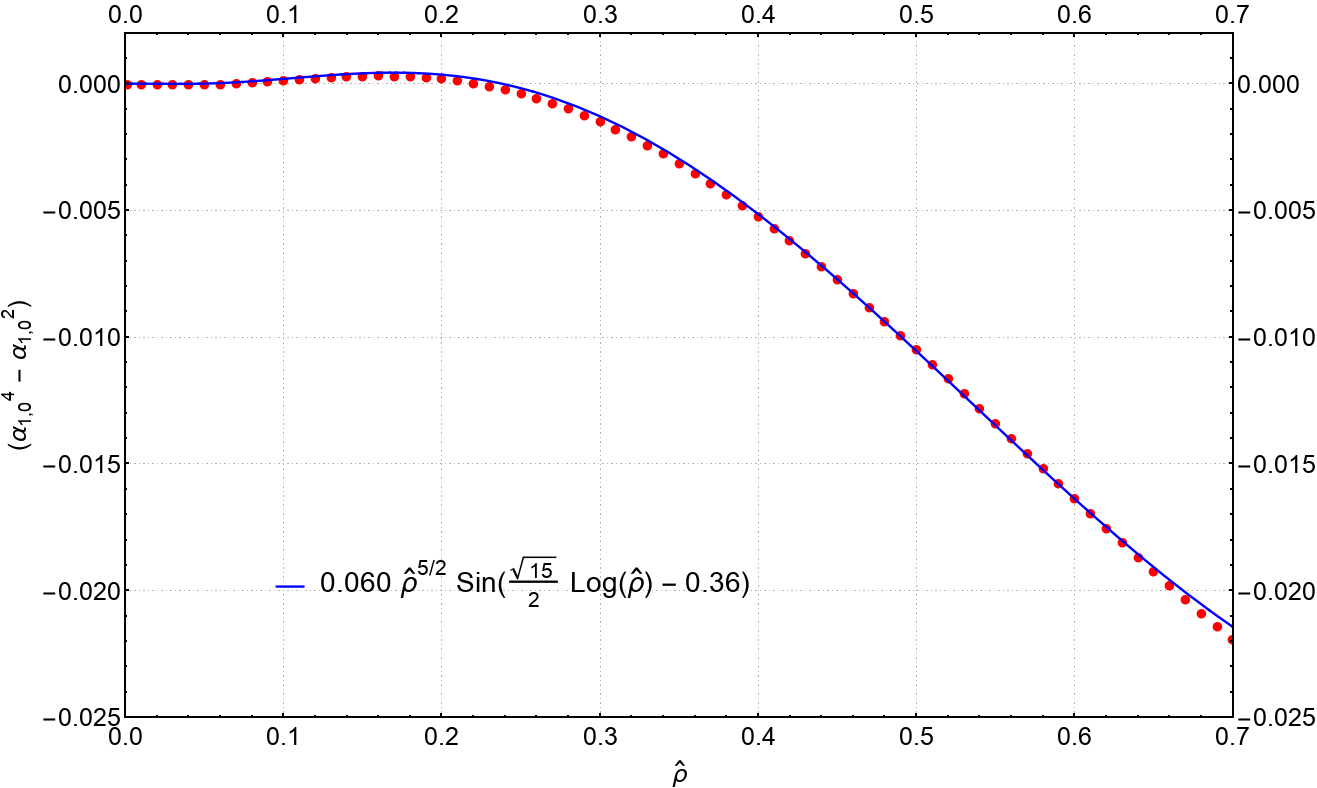} \centering \caption{A fit of $\hat{\alpha}_{1,0}^4 - \hat{\alpha}_{1,0}^2$ for $\mathcal{M} = S^2 \times S^4$ when $S^2$ shrinks.} \label{fig:S2S4coefficientsS2shrinks}
\end{figure}

\section*{Acknowledgements}
We would like to thank M.~Berkooz and M.~Rozali for useful discussions, and to especially thank A.~Buchel and D.~Kutasov for many useful discussions and for comments on a draft of this paper. This work was supported in part  by an Israel Science Foundation center for excellence grant (grant number 1989/14) and by the Minerva foundation with funding from the Federal German Ministry for Education and Research. OA is the Samuel Sebba Professorial Chair of Pure and Applied Physics. 


\end{document}